\documentclass[journal,twocolumn]{IEEEtran}
\IEEEoverridecommandlockouts

\usepackage{algorithm,algorithmic,amssymb,amsthm,bbm,cite,color,enumitem,graphicx,microtype,setspace,tabularray,url}
\usepackage[tbtags]{amsmath}
\usepackage[USenglish]{babel}
\usepackage{hyperref}
\usepackage[utf8]{inputenc}
\usepackage[T1]{fontenc}

\usepackage[labelsep=period,font=small]{caption}
\usepackage[nolist]{acronym}
\usepackage{cite}
\usepackage{placeins}
\usepackage{tikz,pgfplots}
\usetikzlibrary{arrows, arrows.meta, backgrounds, calc, intersections, decorations.markings, decorations.pathreplacing, positioning, shapes}

\usetikzlibrary{external}
\pgfplotsset{compat=1.17,
legend image code/.code={
\draw[mark repeat=2,mark phase=2]
plot coordinates {
(0cm,0cm)
(0.15cm,0cm)        
(0.3cm,0cm)         
};%
}
}

\usepackage{titlesec}
\let\subparagraph\relax
\titlespacing{\section}{0pt}{6pt plus 2pt minus 1pt}{4pt plus 1pt minus 1pt} 
\titlespacing{\subsection}{0pt}{4pt plus 2pt minus 1pt}{2pt plus 1pt minus 1pt} 
\newcommand{\g}{\mathbf{g}}

\renewcommand{\v}{\mathbf{v}}

\newcommand{\0}{\mathbf{0}}

\newcommand{\I}{\mathbf{I}}

\newcommand{\setC}{\mathcal{C}}

\newcommand{\setN}{\mathcal{N}}

\newcommand{\Real}{\mbox{$\mathbb{R}$}}
\newcommand{\Compl}{\mbox{$\mathbb{C}$}}

\newcommand{\diff}{\mathop{}\!\mathrm{d}}
\newcommand{\Exp}{\mathbb{E}}
\newcommand{\herm}{\mathrm{H}}

\mathchardef\mhyphen="2D

\def \am {$\alpha$-$\mu$}
\def \km {$\kappa$-$\mu$}
\def \hm {$\eta$-$\mu$}
\def \Ehm {Extended $\eta$-$\mu$}
\def \ahkm {$\alpha$-$\eta$-$\kappa$-$\mu$}

\definecolor{myred}{HTML}{F05039}
\definecolor{myorange}{HTML}{EEBAB4}
\definecolor{myblue}{HTML}{3D65A5}
\definecolor{mygreen}{HTML}{539054}
\definecolor{myyellow}{HTML}{F0E442}

\let\originalleft\left
\let\originalright\right
\renewcommand{\left}{\mathopen{}\mathclose\bgroup\originalleft}
\renewcommand{\right}{\aftergroup\egroup\originalright}

\begin{acronym}

\acro{3GPP}{the 3rd Generation Partnership Project}
\acro{5G}{fifth-generation}
\acro{6G}{sixth-generation}
\acro{AWGN}{additive white Gaussian noise}
\acro{BEP}{bit error probability}
\acro{BER}{bit error rate}
\acro{BFSK}{binary frequency-shift keying}
\acro{BPSK}{binary phase-shift keying}
\acro{BS}{base station}
\acro{CCDF}{complementary cumulative distribution function}
\acro{CDF}{cumulative distribution function}
\acro{CP}{coverage probability}
\acro{CSI}{channel state information}
\acro{EGC}{equal gain combining}
\acro{FTR}{fluctuating two-ray}
\acro{i.i.d.}{independent and identically distributed}
\acro{LoS}{line-of-sight}
\acro{$M$-PSK}{$M$-ary phase-shift keying}
\acro{$M$-QAM}{$M$-ary quadrature amplitude modulation}
\acro{MGF}{moment-generating function}
\acro{mmWave}{millimeter-wave}
\acro{MRC}{maximum ratio combining}
\acro{MRT}{maximum ratio transmission}
\acro{NLoS}{non-line-of-sight}
\acro{OP}{outage probability}
\acro{PDF}{probability density function}
\acro{RF}{radio frequency}
\acro{RIS}{reflective intelligent surface}
\acro{RV}{random variable}
\acro{SEP}{symbol error probability}
\acro{SER}{symbol error probability}
\acro{SISO}{single-input single-output}
\acro{SNR}{signal-to-noise ratio}
\acro{sub-THz}{sub-terahertz}

 \end{acronym}

\title{Sum of Squared \Ehm{} and \km{} RVs: A New Framework Applied to FR3 and Sub-THz Systems}
\author{Gustavo Rodrigues de Lima Tejerina and Italo Atzeni
\thanks{This work was supported by the Research Council of Finland (336449 Profi6, 348396 HIGH-6G, and 369116 6G~Flagship) and by the European Commission (101095759 Hexa-X-II). \\ \indent The authors are with the Centre for Wireless Communications, University of Oulu, Finland (e-mail: \{gustavo.tejerina, italo.atzeni\}@oulu.fi).}}

\begin{document}

\maketitle

\begin{abstract}
The analysis of systems operating in future frequency ranges calls for a proper statistical channel characterization through generalized fading models. In this paper, we adopt the \Ehm{} and \km{} models to characterize the propagation in FR3 and the sub-THz band, respectively. For these models, we develop a new exact representation of the sum of squared independent and identically distributed random variables, which can be used to express the power of the received signal in multi-antenna systems. Unlike existing ones, the proposed analytical framework is remarkably tractable and computationally efficient, and thus can be conveniently employed to analyze systems with massive antenna arrays. For both the \Ehm{} and \km{} distributions, we derive novel expressions for the probability density function and cumulative distribution function, we analyze their convergence and truncation error, and we discuss the computational complexity and implementation aspects. Moreover, we derive expressions for the outage and coverage probability, bit error probability for coherent binary modulations, and symbol error probability for $M$-ary phase-shift keying and quadrature amplitude modulation. Lastly, we provide an extensive performance evaluation of FR3 and sub-THz systems focusing on a downlink scenario where a single-antenna user is served by a base station employing maximum ratio transmission.
\end{abstract}


\section{Introduction} \label{sec:Intro}

Current \ac{5G} wireless systems support operations in two frequency ranges: sub-6~GHz, referred to in \ac{3GPP} as FR1 and recently expanded up to 7.125~GHz~\cite{3GPP_38.101-1}, and low millimeter-wave, referred to in \ac{3GPP} as FR2 and spanning from 24.25 to 71~GHz~\cite{3GPP_38.101-2}. Future releases of \ac{5G} are expected to extend operations to the upper mid-band, referred to in \ac{3GPP} as FR3 and encompassing the 7.125--24.25~GHz range~\cite{3GPP_38.820}. FR3 offers an attractive balance between bandwidth and coverage, and is characterized by mild propagation conditions compared with FR2~\cite{Kan24}. On the other hand, \ac{6G} wireless systems are envisioned to explore the sub-THz spectrum, ranging from 100 to 300~GHz, to seek extreme bandwidths and accommodate an unprecedented number of users with high data rates~\cite{Raj20}. In general, raising the carrier frequency has a two-fold impact on the propagation. First, the penetration loss and roughness of the materials with respect to the wavelength increase: consequently, the impact of the \ac{NLoS} paths tends to diminish and the \ac{LoS} component becomes dominant. Second, the free-space path loss (for isotropic antennas) increases: thus, more antennas are needed to create physically large arrays that are capable of properly focusing the signal power. These properties are clearly exacerbated when moving up in frequency from FR3 to FR2 and then to the sub-THz~band~\cite{Han22}.

Wireless channels are generally modeled deterministically or statistically~\cite{Ser22}. The deterministic approach captures the channel characteristics through site-specific measurements and simulations using, e.g., ray tracing and finite-difference time-domain. The statistical approach models the fluctuations of the channel through tractable probability distributions to incorporate, e.g., \ac{LoS}, multipath, and scattering components. For example, the classical Rayleigh, Rice, and Nakagami-$m$ models are commonly used to characterize the fading in the sub-6~GHz range. The suitability of these classical fading models to represent millimeter-wave and sub-THz channels has also been studied~\cite{Ye22,Iqb19,Ekt17,Liu24}. However, such models fail to properly capture the intricacies of the propagation at high frequencies, which motivates investigating more generalized ones such as the (Extended) \hm{}, \km, \am{}, \ahkm{}, and \ac{FTR} models. For instance, \cite{Mar21} compared both classical and generalized fading models to measured data at 26, 28, and 39~GHz in \ac{LoS} and \ac{NLoS} scenarios, showing a better fit for the generalized models, i.e, \hm{}, \km{}, \am{}, and \ahkm{}. Similarly,~\cite{Pap21} showed a better goodness of fit for the \am{} model compared with classical models at 143.1~GHz by conducting field trials in both \ac{LoS} and \ac{NLoS} conditions in three different indoor environments. Furthermore,~\cite{Du22,San24} used measurements to demonstrate the aptness of the \ac{FTR} and multi-cluster \ac{FTR} distributions to model the fading at 143.1~GHz. Other works investigated generalized fading models applied to specific scenarios including, e.g., multiple antennas~\cite{Jos22}, reflective surfaces~\cite{Du22,Le24,Pre24}, dual-hop communications~\cite{Li22}, and random atmospheric absorption~\cite{Bha24}.

Statistical channel characterization is constantly evolving to encompass new propagation environments. However, as more intricate scenarios arise, the mathematical complexity of the resulting models increases considerably. For example, when adding multiple antennas or reflective surfaces in the system, expressing the power of the received signal involves the sum or product of \acp{RV}, which greatly complicates the analysis. When successful, these representations are likely to employ elaborate special functions, such as Horn functions and the Fox H-function, as proposed for the \Ehm{} and \km{} models applied to systems with \ac{MRC}~\cite{Bad21,Dix23} and pre-detection \ac{EGC}~\cite{Bad22} at the receiver. However, their widespread use is limited since these special functions are not readily available in common mathematical packages such as MATLAB, Mathematica, and SciPy. To bypass this issue, researchers have resorted to simplified versions of classical and generalized fading models, namely Nakagami-$m$, \Ehm{}, and \km{}, where integer values are assumed for some of the parameters~\cite{Lop17,Hoa20,Tej23}. More recently,~\cite{Alm23a} developed an analytical framework based on complex analysis to represent the exact sum of \ac{i.i.d.} \km{} \acp{RV}. This framework leads to more tractable and computationally efficient expressions (i.e., given in terms of simple Gamma functions) for the \ac{PDF} and \ac{CDF}. These results were exploited to derive expressions for the outage probability and \ac{BEP} with pre-detection \ac{EGC} that also inherit the framework's simplicity. Following a similar approach,~\cite{Alm23b} represented the exact sum of \ac{i.i.d.} \am{} \acp{RV} and obtained the \ac{PDF} and \ac{CDF} for systems with \ac{MRC} and pre-detection \ac{EGC} at the receiver.

\subsection{Contribution}

The analysis of systems operating in future frequency ranges, from FR3 to the sub-THz band, is of paramount importance and calls for a proper statistical channel characterization through generalized fading models. In this paper, we adopt the \Ehm{} and \km{} models to characterize the propagation in FR3 and the sub-THz band, respectively. Indeed, the \Ehm{} and \km{} models provide remarkable flexibility to represent scenarios with dominant \ac{NLoS} and \ac{LoS} components, respectively, which well fit the propagation in FR3 and at sub-THz frequencies, respectively. In multi-antenna systems with \ac{MRT} at the transmitter or \ac{MRC} at the receiver, the power of the received signal can be expressed as a sum of squared RVs. In this regard,~\cite{Bad21} and~\cite{Mil08} proposed expressions for the \ac{PDF} and \ac{CDF} of the sum of squared \ac{i.i.d.} \Ehm{} and \km{} \acp{RV} that employ special functions such as the Horn functions, modified Bessel function, and Marcum Q-function. More importantly, the evaluation of the performance metrics based on these expressions either involves the Fox-H function~\cite{Bad21} and Horn functions~\cite{Dix23}, or requires cumbersome numerical integration. Thus, when raising the carrier frequency, existing analytical frameworks for the considered models become unsuitable to handle massive numbers of antennas due to their inherent computational complexity. To address this issue, we extend the framework in~\cite{Alm23a} and develop a new exact representation of the sum of squared \ac{i.i.d.} \Ehm{} and \km{} \acp{RV}. The resulting framework is remarkably tractable and computationally efficient, and thus can be conveniently employed to analyze FR3 and sub-THz systems with massive antenna arrays.

The contribution of this paper is summarized as follows.
\begin{itemize}
\item[$\bullet$] For both the \Ehm{} and \km{} distributions, we build on the framework based on complex analysis in~\cite{Alm23a} to derive novel expressions for the \ac{PDF} and \ac{CDF} of the sum of squared \ac{i.i.d.} \acp{RV}. The resulting expressions are given in terms of simple Gamma function, which makes them tractable and computationally efficient. Although these expressions include an infinite summation, they can be conveniently truncated with negligible errors even for a limited number of terms. In this context, we analyze the convergence and truncation error. Furthermore, we discuss the computational complexity and implementation aspects.
\item[$\bullet$] Based on the above expressions for the \ac{PDF} and \ac{CDF}, we derive expressions for important performance metrics such as the outage and coverage probability, \ac{BEP} for coherent binary modulations, and \ac{SEP} for \ac{$M$-PSK} and \ac{$M$-QAM}, along with their asymptotic expressions at high \ac{SNR}. In this regard, we provide two distinct approaches with different computational complexities and convergence conditions. The obtained expressions are easy to evaluate numerically and reduce the computation time compared with the solutions in~\cite{Bad21,Dix23}.
\item[$\bullet$] We provide an extensive performance evaluation of FR3 and sub-THz systems under the \Ehm{} and \km{} models, respectively. In this regard, we focus on a downlink system where a single-antenna user is served by a multi-antenna \ac{BS} employing \ac{MRT}, with perfect and imperfect \ac{CSI} at the latter. The evaluation is carried out in terms of distance between the \ac{BS} and the user, number of antennas, transmit power, carrier frequency, fading parameters, and modulation order. 
For example, we observe that an eight-fold increase in the number of antennas enables the carrier frequency to be approximately doubled. Moreover, even in high-frequencies, \ac{LoS}-dominated scenarios, the \ac{NLoS} components continue to make a noticeable contribution to the received signal. Lastly, the \ac{LoS} or \ac{NLoS} path loss exponent has a greater impact than the frequency-dependent path loss factor.

\end{itemize}

\smallskip

\textit{\textbf{Outline.}} The rest of the paper is structured as follows. Section~\ref{sec:newrep} introduces the new representation of the sum of squared \ac{i.i.d.}  \Ehm{} and \km{} \acp{RV}, presenting novel expressions for the \ac{PDF} and \ac{CDF}. Section~\ref{sec:metrics} builds on the proposed framework to derive new expressions for the outage and coverage probability, \ac{BEP}, and \ac{SEP}. Section~\ref{sec:num} provides an extensive performance evaluation of FR3 and sub-THz systems. Lastly, Section~\ref{sec:final} concludes the paper.

\smallskip

\textit{\textbf{Notation.}} Boldface lowercase letters indicate vectors, $\lVert \cdot\rVert$ is the euclidean norm, and $\left(\cdot\right)^{\herm}$ is the Hermitian transpose operator. $\Real$ and $\Compl$ indicate the sets of real and complex numbers, respectively, with $\jmath = \sqrt{-1}$. $\mathbb{E}[\cdot]$ is the expectation operator. $f_X(x)$, $F_X(x)$, and $\mathcal{M}_X(s)$ correspond to the \ac{PDF}, \ac{CDF}, and \ac{MGF}, respectively, of \ac{RV} $X$.
$\text{erfc}(\cdot)$ denotes the complementary error function \cite[Eq.~(06.27.07.0001.01)]{WolframResearch}, $\Gamma(\cdot)$ is the Gamma function~\cite[Eq.~(06.05.02.0001.01)]{WolframResearch}, $\Gamma(\cdot,\cdot)$ is the incomplete Gamma function~\cite[Eq.~(06.06.02.0001.01)]{WolframResearch}, and $B(\cdot;\cdot,\cdot)$ indicates the incomplete Beta function~\cite[Eq.~(06.19.02.0001.01)]{WolframResearch}. $I_\nu(\cdot)$ is the modified Bessel function of the first kind and $\nu$-th order~\cite[Eq.~(03.02.02.0001.01)]{WolframResearch}. $_1 F_1 (\cdot;\cdot;\cdot)$ is the Kummer confluent hypergeometric function~\cite[Eq.~(07.20.02.0001.01)]{WolframResearch}. $ _2F_1(\cdot,\cdot;\cdot;z)$ for $|z|<1$ denotes the Gauss hypergeometric function~\cite[Eq.~(07.23.02.0001.01)]{WolframResearch}. $_p F_q (a_1,\ldots,a_p;b_1,\ldots,b_q;z)$ is the generalized hypergeometric function~\cite[Eq.~(07.31.02.0001.01)]{WolframResearch} for $|z|<1$ if $q=p-1$ or $\forall z \in \Real$ if $q \geq p $. $F_1 (\cdot;\cdot,\cdot;\cdot;z_1,z_2)$ for $|z_1|<1$ and $|z_2|<1$ is the Appell hypergeometric function~\cite[Eq.~(07.36.02.0001.01)]{WolframResearch}. Lastly, $\sim$ means asymptotically equivalent.

\section{Sum of Squared \acp{RV}: New Framework}\label{sec:newrep}

Considering a multi-antenna \ac{BS} serving a single-antenna user, the sum of squared \acp{RV} can be used to express the instantaneous uplink \ac{SNR} when the \ac{BS} adopts \ac{MRC} or post-detection \ac{EGC}, or the instantaneous downlink \ac{SNR} when the \ac{BS} adopts \ac{MRT}. In this paper, we focus on the latter as our motivating scenario and for our performance evaluation, although the proposed analytical framework also applies to the other settings.

In this context, consider a \ac{BS} equipped with $N$ antennas transmitting data to a single-antenna user. The signal received at the user is given by
\begin{align}
y = \sqrt{P_{\textrm{t}}} \g^{\herm} \v x + z \in \Compl,
\end{align}
where $P_{\textrm{t}}$ is the transmit power, $\g = [g_{1}, \ldots, g_{N}] \in \Compl^{N}$ is the channel vector, $\v \in \Compl^{N}$ is the precoding vector (with $\| \v \|^{2} = 1$), $x \in \Compl$ is the transmitted data symbol (with $\Exp\big[ |x|^{2} \big] = 1$), and $z \sim \setC \setN(0, \sigma^{2})$ is the noise term. With perfect \ac{CSI}, the \ac{MRT} precoding vector is given by $\v = \g/\| \g \|$ and the corresponding instantaneous \ac{SNR} at the user is
\begin{align} \label{eq:perfectSNR}
\textrm{SNR}_{\textrm{P}} = \frac{P_{\textrm{t}}}{\sigma^{2}} |\g^{\herm} \v|^{2} = \frac{P_{\textrm{t}}}{\sigma^{2}} \sum_{n=1}^{N} |g_{n}|^{2}.
\end{align}
Alternatively, assuming a simplified model for imperfect \ac{CSI}, the \ac{MRT} precoding vector is given by $\v = \hat{\g}/\|\hat{\g}\|$, where $\hat{\g} = \sqrt{1-\alpha^{2}} \g + \alpha\tilde{\g} \in \Compl^N$ denotes the estimated channel, $\tilde{\g} \sim \setC \setN \big(\0, \frac{1}{N} \| \g \|^{2} \I_{N}\big)$ is the channel estimation error (independent of $\g$), and $\alpha \in [0,1]$ is the estimation accuracy (with $\alpha=0$ representing perfect \ac{CSI}). When $N \gg 1$, the corresponding instantaneous \ac{SNR} at the user can be approximated as
\begin{align} \label{eq:imperfectSNR}
\textrm{SNR}_{\textrm{I}} \simeq (1-\alpha^{2}) \textrm{SNR}_{\textrm{P}},
\end{align}
with $\textrm{SNR}_{\textrm{P}}$ defined in \eqref{eq:perfectSNR}. Note that \eqref{eq:perfectSNR} and \eqref{eq:imperfectSNR} can also be used to characterize the instantaneous uplink \ac{SNR} when the \ac{BS} adopts \ac{MRC}.

Motivated by the above system model, we propose a new exact representation of the sum of squared \ac{i.i.d.} \Ehm{} and \km{} \acp{RV}. The proposed framework is based on the recursive solution provided in~\cite{Alm23a} and results in much simpler expressions for the \ac{PDF} and \ac{CDF} (i.e., given in terms of simple Gamma functions) compared with~\cite{Mil08,Bad21,Dix23}, which employed special functions such as the modified Bessel function, Marcum Q-function, and Horn functions. Hence, the resulting performance metrics, such as the \ac{BEP} and \ac{SEP}, also inherit the framework's simplicity. Furthermore, we analyze the convergence, truncation error, and computational complexity of these new expressions. The proposed framework has proven to return fast and precise solutions even when employing a very large number of antennas. In the rest of this section (as well as in Section~\ref{sec:metrics}), we present all the detailed steps for the \Ehm{} distribution and only a sketch of the procedure for the \km{} distribution, which follows the same rationale.

\subsection{\ac{PDF} and \ac{CDF} for the \Ehm{} Distribution} \label{sec:newrep_Ehm}

The \Ehm{} model is a generalized fading model used to represent scenarios with dominant \ac{NLoS} components with power and cluster imbalances~\cite{Tej20}. These features are enabled through the parameters $\eta$, $\mu$, and $p$, where $\eta > 0$ denotes the power ratio between the in-phase and quadrature signals, $\mu > 0$ is the number of multipath clusters, and $p > 0$ represents the ratio between the numbers of multipath clusters of the in-phase and of the quadrature components. The \Ehm{} model is known to encompass classical fading models such as Hoyt, Nakagami-$m$, and Rayleigh. For the \Ehm{} distribution, the \ac{PDF} of the envelope of the $n$-th \ac{RV}, denoted by $R_{n} = \sqrt{\frac{P_{\textrm{t}}}{\sigma^{2}}} |g_{n}|$, is defined as~\cite[Eq.~(14)]{Tej20}
\begin{align} \label{eq:hm:fr}
f_{R_n}(r_n) = \frac{2\xi^{\mu}r_n^{2 \mu -1} \,_1F_1\big(\frac{\mu  p}{1+p};\mu ;\frac{\xi  (\eta -p) r_n^2}{\eta \hat{w}_n}\big)}{\Gamma (\mu ) \hat{w}_n^{\mu} \exp\big(\xi\frac{r_n^2}{\hat{w}_n}\big) } \left(\frac{p}{\eta }\right)^{\frac{\mu  p}{1+p}},
\end{align}
with $\xi=\mu \left(1 + \eta\right)/ \left( 1+p\right)$, and where $\hat{w}_n = \mathbb{E}[R_n^2]$ represents the expectation of the squared envelope of the $n$-th RV.
For the new representation of the sum of $N$ squared \ac{i.i.d.} \acp{RV}, we follow the standard \ac{MGF}-based approach to obtain the \ac{PDF}. This procedure consists in first deriving the \ac{MGF} of a transformed \ac{RV} (i.e., the square of the original \ac{RV}) and then expressing the sum as the product of $N$ \acp{MGF}.

For the first step, we begin by applying the \ac{RV} transformation $W_n = R_n^2 = \frac{P_{\textrm{t}}}{\sigma^{2}} |g_{n}|^{2}$ in \eqref{eq:hm:fr}, resulting in
\begin{align} \label{eq:hm:fwm}
   f_{W_n}(w_n) = \frac{\xi^{\mu}w_n^{\mu -1} \, _1F_1 \big(\frac{\mu  p}{1+p};\mu ;\frac{\xi (\eta -p) w_n}{\eta  \hat{w}_n }\big)}{\Gamma (\mu ) \hat{w}_n ^{\mu } \exp \big(\xi\frac{w_n}{ \hat{w}_n }\big)}  \left(\frac{p}{\eta }\right)^{\frac{\mu  p}{1+p}}.
\end{align}
Then, the \ac{MGF} of \eqref{eq:hm:fwm} is obtained after applying the Laplace transform, i.e., $ \mathcal{M}_{X}(s) = \int_0^\infty f_X(x)\exp\left( -s x \right)\diff x$, leading to
\begin{align}
    \mathcal{M}_{W_n}(s) = \ & \frac{\xi^{\mu}}{\Gamma (\mu ) \hat{w}^{\mu}}  \left(\frac{p}{\eta }\right)^{\frac{\mu  p}{1+p}} \int_0^{\infty} \frac{w_n^{\mu -1}}{\exp \big(s w_n+ \frac{\xi w_n}{ \hat{w}_n }\big)} \nonumber \\
    &\times \, _1F_1\left(\frac{\mu  p}{1+p};\mu ;\frac{\xi (\eta -p) w_n}{\eta  \hat{w}_n }\right) \diff w_n. \label{eq:eta:mgf1}
\end{align}
Following the same rationale of~\cite{Alm23a}, we rewrite the exponential function and Kummer confluent hypergeometric function $_1F_1$ according to their contour integral representations~\cite[Eq.~(07.20.07.0003.01), (01.03.07.0001.01)]{WolframResearch}, which results in
\begin{align}
        \mathcal{M}_{W_n}(s) = \ & \frac{\xi^{\mu} \big(\frac{p}{\eta }\big)^{\frac{\mu  p}{1+p}}}{(2 \pi  \jmath)^{2} \Gamma \big(\frac{\mu  p}{1+p}\big)} \int_{0}^{\infty} \oint_{\mathcal{L}_t} \oint_{\mathcal{L}_v} \frac{\Gamma (t) \Gamma (v) \big(\frac{\eta  \xi }{p-\eta }\big)^v}{\Gamma (\mu -v)} \nonumber \\
        & \times \frac{\Gamma \big(\frac{\mu  p}{1+p}-v\big)  w^{\mu -t-v-1}_n}{\exp \big(\frac{\xi w_n}{\hat{w}_n }\big) s^{t} \hat{w}_n ^{\mu -v}} \diff v \diff t \diff w_n, \label{eq:eta:mgf-contour1}
\end{align}
where $\mathcal{L}_t$ and $\mathcal{L}_v$ are the complex contours of $t$ and $v$, respectively. After this procedure, the integral operators in \eqref{eq:eta:mgf-contour1} are reordered to solve it in terms of $w_n$, leading to 
\begin{align}
    \mathcal{M}_{W_n}(s) = \ & \frac{\left(2 \pi \jmath\right)^{-2}}{\Gamma \big(\frac{\mu  p}{1+p}\big)}  \bigg(\frac{p}{\eta }\bigg)^{\frac{\mu  p}{1+p}} \oint_{\mathcal{L}_t^{*}} \oint_{\mathcal{L}_v^{*}} \frac{\Gamma (t) \Gamma (v)}{\Gamma (\mu -v)} \nonumber \\
        &\times \frac{\Gamma (\mu -t-v) \Gamma \big(\frac{\mu p}{1+p}-v\big)}{\big(\frac{p}{\eta }-1\big)^{v} \big(\frac{\xi }{ s \hat{w}_n }\big)^{-t}} \diff v \diff t, \label{eq:eta:mgf-contour2}
\end{align}
where $\mathcal{L}_t^{*}$ and $\mathcal{L}_v^{*}$ are the new complex contours of $t$ and $v$, respectively, after the integration over $w_n$. With this procedure, \eqref{eq:eta:mgf-contour2} has new essential singularities located at $t \in \{-m, (l-m+\mu)\}$ and at $v \in \{l+\mu p/(1+p),l+\mu,-l \}$. As in~\cite{Alm23a}, by identifying the singularities, the new complex contours are defined to guarantee the convergence of \eqref{eq:eta:mgf-contour2} and circumvent duplicate poles. With this in mind, both contour integrals can be solved by applying the residue theorem~\cite[Eq.~(16.3.6)]{Kre10} to \eqref{eq:eta:mgf-contour2} for the poles $t \to m-l+\mu$ and $v\to -l$, obtaining
\begin{align}
       \mathcal{M}_{W_n}(s) = \ & \frac{\big(\frac{p}{\eta }\big)^{\frac{\mu  p}{1+p}}}{\Gamma \big(\frac{\mu  p}{1+p}\big)}    \sum _{m=0}^{\infty } \sum _{l=0}^{\infty } \left(\frac{p}{\eta }-1\right)^l \left(\frac{\xi }{s \hat{w}_n }\right)^{l+m+\mu } \nonumber \\
       & \times \frac{\Gamma (l+m+\mu ) \Gamma \big(l+\frac{\mu p }{1+p}\big)}{(-1)^m (-1)^l m! l! \Gamma (l+\mu )} . \label{eq:eta:mgf-sum1}
\end{align}
Finally, \eqref{eq:eta:mgf-sum1} is rearranged in terms of Cauchy product as
\begin{align}
        \mathcal{M}_{W_n}(s) = \ & \frac{\big(\frac{p}{\eta }\big)^{\frac{\mu  p}{1+p}} \big(\frac{\xi}{ s \hat{w}_n }\big)^{\mu }}{\Gamma \big(\frac{\mu  p}{1+p}\big)}   \sum _{m=0}^{\infty } \left(-\frac{\xi}{ s \hat{w}_n}\right)^m  \sum _{l=0}^m \frac{\Gamma (\mu +m)}{(m-l)!}\nonumber \\
& \times \frac{\Gamma \big(-l+m+\frac{\mu p }{1+p}\big)\big(\frac{p}{\eta }-1\big)^{m-l}}{l! \Gamma (-l+m+\mu )}. \label{eq:eta:mgf-sum2}
\end{align}

For the second step, the \ac{MGF} of the sum of $N$ \ac{i.i.d.} \acp{RV} $W_{n}$, denoted by $W = \sum_{n=1}^{N} W_{n} = \textrm{SNR}_{\textrm{P}}$, is obtained as the product of their respective \acp{MGF}, i.e., $\mathcal{M}_{W} (s) = \prod_{n=1}^{N} \mathcal{M}_{W_n} (s)$. Hence, from \eqref{eq:eta:mgf-sum2} and~\cite[Eq.~(0.314)]{Gra07}, the product of the \acp{MGF} is given by
\begin{align} \label{eq:eta:mgf_sum}
        \mathcal{M}_{W}(s) = \Bigg( \frac{\xi^{\mu} \big(\frac{p}{\eta }\big)^{\frac{\mu p}{1+p}}}{\Gamma \big(\frac{\mu  p}{1+p}\big)} \Bigg)^{N} \sum_{m=0}^{\infty}\bigg(\frac{1}{s \hat{w}}\bigg)^{N\mu +m} h_m,
\end{align}
where we have replaced $\hat{w}_{n} = \hat{w}$, $\forall n$ due to the \ac{i.i.d.} property and with
\begin{subequations} \label{eq:eta:hm}
\begin{align}
    h_0 = \ & \Gamma \bigg(\frac{\mu  p}{1+p}\bigg)^N, \label{eq:eta:hm0} \\
    h_m = \ & \sum _{i=1}^m  \frac{ (i N+i-m)  h_{m-i}}{m \Gamma \big(\frac{\mu  p}{1+p}\big) \left(-\xi\right)^{-i}} \sum _{l=0}^i\frac{ \Gamma(i+\mu)}{l! (i-l)! } \nonumber \\
    & \times \frac{ \Gamma \big(i-l+\frac{\mu p }{1+p}\big) \big(\frac{p}{\eta }-1\big)^{i-l}}{ \Gamma (i-l+\mu )}, \, m\geq 1. \label{eq:eta:hm1}
\end{align}
\end{subequations}
Finally, the inverse Laplace transform is applied to \eqref{eq:eta:mgf_sum} by using~\cite[Eq.~(17.13.3)]{Gra07} and the \ac{PDF} is obtained as
\begin{align} \label{eq:eta:pdf_snr}
        f_{W}(w) = \Bigg(\frac{\big(\frac{p}{\eta }\big)^{\frac{\mu  p}{1+p}} \xi^{\mu }}{\Gamma \big(\frac{\mu  p}{1+p}\big)}\Bigg)^N \sum _{m=0}^{\infty } \frac{ w^{N \mu + m-1} h_m}{\hat{w} ^{N \mu + m } \Gamma (N \mu + m )},
\end{align}
with $h_m$ defined in \eqref{eq:eta:hm}. The \ac{CDF} readily follows from applying the standard procedure, i.e., $F_X(x) = \int_0^{x} f_X(x)\diff x $, which yields
\begin{align} \label{eq:eta:cdf_snr}
        F_{W}(w) = \Bigg(\frac{\big(\frac{p}{\eta }\big)^{\frac{\mu  p}{1+p}} \xi^{\mu }}{\Gamma \big(\frac{\mu  p}{1+p}\big)}\Bigg)^N \sum _{m=0}^{\infty } \frac{\big(\frac{w}{\hat{w}}\big)^{m+\mu  N} h_m }{\Gamma (N \mu + m +1)}.
\end{align}

An alternative \ac{MGF} representation can be obtained from \eqref{eq:eta:mgf-sum1} by simplifying its $m$-indexed summation and reducing it to a power function, leading to 
\begin{align}
    \mathcal{M}_{W{n}}(s)= \ & \frac{\big(\frac{p}{\eta }\big)^{\frac{\mu  p}{1+p}}}{\Gamma \big(\frac{\mu  p}{1+p}\big)} \sum _{l=0}^{\infty } \frac{\big(1 \! - \! \frac{p}{\eta }\big)^l \Gamma \big(l \! + \! \frac{\mu  p}{1+p}\big)}{\Gamma (l+1)} \Bigg(\frac{1 }{1 \! + \! \frac{s \hat{w}}{\xi} }\Bigg)^{l+\mu }. \label{eq:eta:mgf-pb1}
\end{align}
After rearranging \eqref{eq:eta:mgf-pb1} and with the help of~\cite[Eq.~(0.314)]{Gra07}, we obtain the reformulated \ac{MGF} as (cf. \eqref{eq:eta:mgf_sum})
\begin{align} \label{eq:eta:mgf-pb2}
        \mathcal{M}_{W}(s)&= \ \Bigg(\frac{\big(\frac{p}{\eta }\big)^{\frac{\mu  p}{1+p}}}{\Gamma \big(\frac{\mu  p}{1+p}\big)}\Bigg)^{N} \sum _{m=0}^{\infty } \Bigg(\frac{1 }{1 +\frac{s \hat{w}}{\xi} }\Bigg)^{N\mu +m } \tilde{h}_m,
\end{align}
with (cf. \eqref{eq:eta:hm})
\begin{subequations} \label{eq:eta:hm:pb}
\begin{align}
    \tilde{h}_0 = \ & \Gamma \bigg(\frac{\mu  p}{1+p}\bigg)^N, \label{eq:eta:hm0:pb} \\
    \tilde{h}_m = \ & \frac{1}{m \Gamma \big(\frac{\mu  p}{1+p}\big)} \sum _{k=1}^m (k N+k-m) \tilde{h}_{m-k} \nonumber \\
    & \times \frac{\big(1-\frac{p}{\eta }\big)^k \Gamma \big(k+\frac{\mu p }{1+p}\big)}{\Gamma (k+1)}, \, m\geq 1. \label{eq:eta:hm1:pb}
\end{align}
\end{subequations}
From \eqref{eq:eta:mgf-pb2}, the PDF is obtained as (cf. \eqref{eq:eta:pdf_snr})
\begin{align} \label{eq:eta:pdf_snr2}
        f_{W}(w) = \Bigg(\frac{\big(\frac{p}{\eta }\big)^{\frac{\mu  p}{1+p}}}{\Gamma \big(\frac{\mu  p}{1+p}\big)}\Bigg)^{N}  \sum_{m=0}^{\infty}\frac{ \big(\frac{\xi w}{\hat{w}}\big)^{N \mu + m} \tilde{h}_m}{ w \Gamma (N \mu + m ) \exp\big(\frac{\xi w}{\hat{w}}\big)},
\end{align}
whereas the CDF is given by (cf. \eqref{eq:eta:cdf_snr})
\begin{align} \label{eq:eta:cdf_snr2}
        F_{W}(w) = \Bigg(\frac{\big(\frac{p}{\eta }\big)^{\frac{\mu  p}{1+p}}}{\Gamma \big(\frac{\mu  p}{1+p}\big)}\Bigg)^{N}\sum_{m=0}^{\infty}\Bigg(1-\frac{ \Gamma \big(N \mu + m,\frac{ \xi w}{\hat{w} }\big)}{ \Gamma (N \mu + m )}\Bigg)\tilde{h}_m.
\end{align}

In Section~\ref{sec:metrics}, we will use the \acp{PDF} in \eqref{eq:eta:pdf_snr} and \eqref{eq:eta:pdf_snr2} (resp. the \acp{MGF} in \eqref{eq:eta:mgf_sum} and \eqref{eq:eta:mgf-pb2}) to derive two distinct sets of expressions for the \ac{BEP} (resp. \ac{SEP}): one with lower computational complexity but quite restrictive convergence conditions, and another with slightly higher computational complexity and no restrictions on the convergence. The same applies to the \acp{PDF} in \eqref{eq:pdf_snr} and \eqref{eq:kappa:pdf_snr2} (resp. \acp{MGF} in \eqref{eq:kappa:mgf} and \eqref{eq:km:mgf-pb2}) presented in the next section for the \km{} distribution.

\subsection{\ac{PDF} and \ac{CDF} for the \km{} Distribution} \label{sec:newrep_km}

The \km{} model is a generalized fading model used to represent scenarios with a dominant \ac{LoS} component~\cite{Yac07}. This feature is enabled through the parameters $\kappa$ and $\mu$, where $\kappa > 0$ represents the power ratio between the \ac{LoS} and \ac{NLoS} (i.e., scattered) components and $\mu > 0$ indicates the number of multipath clusters (as in the \Ehm{} model). The \km{} model is known to encompass classical fading models such as Nakagami-$m$, Rice, and Rayleigh. For the \km{} distribution, the \ac{PDF} of the envelope of the $n$-th \ac{RV} is defined as~\cite[Eq.~(1)]{Yac07}
\begin{align} \label{eq:pdfkappa}
    f_{R_n}(r_n) = \frac{2 \mu  (\kappa +1)^{\frac{\mu +1}{2}} r_n^{\mu} I_{\mu -1}\big(2 \mu  \sqrt{\kappa  (\kappa +1) \frac{r_{n}^{2}}{\hat{w}_{n}}}\big)}{\kappa ^{\frac{\mu -1}{2}} \hat{w}_{n}^{\frac{\mu +1}{2}} \exp \big(\kappa  \mu +(\kappa +1) \mu  \frac{r_{n}^{2}}{\hat{w}_{n}}\big)}.
\end{align}
For the new representation of the sum of $N$ squared \ac{i.i.d.} \km{} \acp{RV}, we follow similar steps as for the \Ehm{} distribution.

For the first step, we begin by applying the \ac{RV} transformation $W_n = R_n^2$. Then, we apply the Laplace transform and use the contour representations of the exponential and Bessel functions~\cite[Eq.~(03.02.07.0009.01)]{WolframResearch}. Furthermore, the contour integrals are solved for the poles $t\to \frac{1}{2} (\mu +2 m-2 v+1)$ and $v\to \frac{1}{2} (-2 l-\mu +1)$, which results in the \ac{MGF} of $W_n$. For the second step, we obtain the product of the \acp{MGF} as
\begin{align} \label{eq:kappa:mgf}
        \mathcal{M}_W (s) = \left(\frac{(\kappa +1) \mu }{\exp(\kappa) }\right)^{N \mu } \sum_{m=0}^{\infty}\left(\frac{1}{s \hat{w}}\right)^{N \mu + m} k_m,
\end{align}
with
\begin{subequations} \label{eq:km}
\begin{align}
    k_0 = \ & 1, \label{eq:km0} \\
    k_m = \ & \frac{1}{m} \sum_{i=1}^{m}(i-m+N i) \Gamma (i+\mu ) k_{m-i} \nonumber \\
    & \times \sum_{l=0}^i \frac{\left(-\kappa  \mu \right)^{-l} \left(\kappa  (\kappa +1) \mu ^2\right)^i}{l! (i-l)! \Gamma (-l+i+\mu )}, \, m\geq 1. \label{eq:km1}
\end{align}
\end{subequations}
Finally, the inverse Laplace transform is applied to \eqref{eq:kappa:mgf} and the \ac{PDF} is obtained as
\begin{align} \label{eq:pdf_snr}
        f_{W}(w) = \left(\frac{(\kappa +1) \mu }{\exp(\kappa) }\right)^{N \mu } \sum_{m=0}^{\infty}\frac{ w^{N \mu + m-1} k_m}{ \hat{w}^{N \mu+ m} \Gamma (N \mu + m )},
\end{align}
with $k_m$ defined in \eqref{eq:km}. The \ac{CDF} readily follows as
\begin{align} \label{eq:cdf_snr}
        F_{W}(w) = \left(\frac{(\kappa +1) \mu }{\exp(\kappa) }\right)^{N \mu } \sum_{m=0}^{\infty}\frac{\big(\frac{w}{\hat{w}}\big)^{N \mu + m} k_m}{\Gamma (N \mu + m + 1)}.
\end{align}

An alternative \ac{MGF} representation can be obtained  as (cf. \eqref{eq:kappa:mgf})
\begin{align} \label{eq:km:mgf-pb2}
        \mathcal{M}_{W}(s)&=  \sum_{m=0}^{\infty} \frac{\tilde{k}_m}{\exp(N \kappa \mu)} \bigg(\frac{  K }{ K + s \hat{w} }\bigg)^{N\mu +m} ,
\end{align}
with (cf. \eqref{eq:km})
\begin{subequations} \label{eq:km:pb-km}
\begin{align}
    \tilde{k}_0 = \ & 1, \label{eq:km:pb-km-0} \\
    \tilde{k}_m = \ & \frac{1}{m} \sum _{i=1}^m \frac{(N i +i-m)(\kappa \mu )^i \tilde{k}_{m-i}}{\Gamma(i+1)}, \, m\geq 1. \label{eq:km:pb-km-m}
\end{align}
\end{subequations}
From \eqref{eq:km:mgf-pb2}, the PDF is obtained as (cf. \eqref{eq:pdf_snr})
\begin{align} \label{eq:kappa:pdf_snr2}
        f_{W}(w) = \frac{\exp\big(\!-\frac{K w}{\hat{w}}\big)}{\exp(N\kappa\mu) } \sum_{m=0}^{\infty}\frac{ \big(\frac{K w}{\hat{w}}\big)^{N \mu + m} \tilde{k}_m}{ w \Gamma (N \mu + m )},
\end{align}
whereas the CDF is given by (cf. \eqref{eq:cdf_snr})
\begin{align} \label{eq:kappa:cdf_snr2}
        F_{W}(w) = \sum_{m=0}^{\infty}\Bigg(1-\frac{ \Gamma \big(N \mu + m,\frac{ K w}{\hat{w} }\big)}{ \Gamma (N \mu + m )}\Bigg)\frac{\tilde{k}_m}{\exp(N\kappa\mu)}.
\end{align}

\subsection{Convergence and Truncation Analysis} \label{sec:trunc}

In this section, we analyze the convergence of the infinite summations in \eqref{eq:eta:pdf_snr}--\eqref{eq:eta:cdf_snr} and \eqref{eq:pdf_snr}--\eqref{eq:cdf_snr}. In calculus, the convergence of a numerical series is commonly inferred through its absolute convergence, which occurs when the sum of the absolute values of the individual terms converges~\cite[Eq.~(0.21)]{Gra07}. Hence, we prove the absolute convergence by obtaining an upper bound on the summations in \eqref{eq:eta:pdf_snr}--\eqref{eq:eta:cdf_snr} and \eqref{eq:pdf_snr}--\eqref{eq:cdf_snr}, as done in~\cite{Alm23a}. With this outcome, we derive the truncation error associated with the \ac{PDF} and \ac{CDF} of each distribution, which provides useful insights into the number of terms needed to numerically compute these functions with the desired accuracy.

For the summations in \eqref{eq:eta:pdf_snr}--\eqref{eq:eta:cdf_snr} of the \Ehm{} distribution, we need to ensure that the inequality
\begin{align} \label{eq:eta:ub}
   \mathcal{C}_{\eta} = \sum_{m=0}^{\infty}\frac{w^{N \mu + m-1+\zeta}}{\hat{w} ^{N \mu + m}\Gamma (N \mu + m +\zeta)} \left|h_m\right| < \infty
\end{align}
holds, where $\zeta = 0$ and $\zeta = 1$ correspond to \eqref{eq:eta:pdf_snr} and \eqref{eq:eta:cdf_snr}, respectively. To this end, we first upper bound $\left|h_m\right|$ for $m\geq 1$ as
\begin{align} \label{eq:eta:up1}
    \left| h_m\right| < \ & \frac{1}{m \Gamma \big(\frac{\mu  p}{1+p}\big)} \sum _{i=1}^m (N i +i-m) \Gamma(i+\mu) \xi^i \left|h_{m-i}\right| \nonumber \\
    & \times \sum _{l=0}^i\frac{\Gamma \big(i-l+\frac{\mu p }{1+p}\big)}{l! (i-l)! \Gamma (i-l+\mu )} \left(\frac{p}{\eta }-1\right)^{i-l}.
\end{align}
Since $\Gamma\big(i-l+\mu p/(1+p)\big)/\Gamma(i-l+\mu) \leq \Gamma\big(\mu p/(1+p)\big)/\Gamma(\mu)$ and $1/\Gamma(\mu) < 8/7$ for any choice of $\mu$ and $p$, we can upper bound the $l$-indexed summation in \eqref{eq:eta:up1} by a power function~\cite[Eq.~(1.111)]{Gra07}, leading to  
\begin{align} \label{eq:eta:up2}
    \left| h_m\right| < \frac{8}{7 m} \sum _{i=1}^m  \frac{ (N i+i-m)\Gamma\left(i+
    \mu\right) \left|h_{m-i}\right|}{\Gamma(i+1) \big(\frac{\xi p}{\eta}\big)^{-i}}. 
\end{align}
Considering the last term of the summation (corresponding to $i = m$) and the fact that $1/\Gamma(m+1) < 1/\Gamma(m)$, we obtain the simpler upper bound 
\begin{align} \label{eq:eta:up3}
    | h_m| < \frac{8 N \Gamma(m+\mu)}{7 \Gamma(m)} \left(\xi\frac{p}{\eta}\right)^m \left|h_{0}\right|.
\end{align}
Now, we extend \eqref{eq:eta:up3} to all $m$ by substituting the right-hand side of \eqref{eq:eta:up3} into \eqref{eq:eta:ub} and performing the index change $m = k + 1$. After some mathematical manipulations, we obtain the following upper bound on \eqref{eq:eta:ub}:
\begin{align}
    \mathcal{C}_{\eta} < \ & 
    \frac{w^{N \mu-1+\zeta} |h_0| }{\hat{w} ^{N \mu}\Gamma( N\mu+\zeta)} \bigg(1 +\frac{8 N \Gamma (N\mu+\zeta) \Gamma (\mu +1)}{7 \Gamma (N \mu +1+\zeta)} \nonumber \\
    & \times \frac{w \xi  p}{\hat{w} \eta } \, _1F_1\left(\mu +1;N \mu+1+\zeta;\frac{w \xi  p}{\hat{w} \eta }\right)\bigg). \label{eq:eta:ub3}
\end{align}
Finally, the absolute convergence of \eqref{eq:eta:ub} and, consequently, the convergence of \eqref{eq:eta:pdf_snr}--\eqref{eq:eta:cdf_snr}, follows from observing that the Kummer confluent hypergeometric function $_1F_1$ in \eqref{eq:eta:ub3} is finite for any choice of parameters in our framework.

Furthermore, we derive an upper bound on the truncation error when applied to \eqref{eq:eta:pdf_snr}--\eqref{eq:eta:cdf_snr}. We begin by defining the truncation error as the sum of the truncated terms, i.e.,
\begin{align} \label{eq:eta:pdferror}
        \mathcal{E}_{\eta}(\epsilon) = \Bigg(\frac{\big(\frac{p}{\eta }\big)^{\frac{\mu  p}{1+p}} \xi^{\mu }}{\Gamma \big(\frac{\mu  p}{1+p}\big)}\Bigg)^N \sum _{m=\epsilon}^{\infty } \frac{\big(\frac{w}{\hat{w}}\big)^{N \mu+m} h_m}{w ^{1-\zeta } \Gamma (N \mu + m +\zeta )},
\end{align}
where $\epsilon$ indicates the number of terms employed to evaluate \eqref{eq:eta:pdf_snr}--\eqref{eq:eta:cdf_snr}. From \eqref{eq:eta:pdferror}, we follow similar steps as those leading to \eqref{eq:eta:ub3} and, after some mathematical manipulations, we obtain the upper bound
\begin{align}
   \mathcal{E}_{\eta}(\epsilon) < \ &   \frac{8 N \Gamma (\mu +\epsilon)\xi^{N\mu+\epsilon} w^{N \mu+\epsilon-1+\zeta} }{7 \Gamma (\epsilon) \Gamma (N \mu +\epsilon + \zeta)\hat{w} ^{N \mu+\epsilon}} \bigg(\frac{p}{\eta }\bigg)^{\frac{N \mu p}{1+p}+\epsilon}  \nonumber \\
    & \times \, _2F_2\left(1,\mu +\epsilon;N \mu+\epsilon+\zeta,\epsilon;\frac{\xi p w}{  \eta \hat{w}}\right). \label{eq:eta:truncerror}
\end{align}

For the summations in \eqref{eq:pdf_snr}--\eqref{eq:cdf_snr} of the \km{} distribution, we follow similar steps as for the summations in \eqref{eq:eta:pdf_snr}--\eqref{eq:eta:cdf_snr} of the \Ehm{} distribution. Hence, we obtain the upper bound 
\begin{align}
    \mathcal{C}_{\kappa} < \ & \frac{w^{N \mu-1+\zeta}}{\hat{w} ^{N \mu} \Gamma (N \mu +\zeta )} \bigg( 1 +\frac{8 N  \Gamma (\mu +1) \Gamma (N \mu +\zeta) w}{7 \Gamma (N \mu +1 + \zeta) \hat{w}} \nonumber \\
    & \times K \, _1F_1\bigg(\mu +1;N \mu+1+\zeta;\frac{K w}{ \hat{w}}\bigg)\bigg), \label{eq:km:ub}
\end{align}
with $K = \mu (\kappa+1) (\kappa\mu+1)$, and where $\zeta = 0$ and $\zeta = 1$ correspond to  \eqref{eq:pdf_snr} and \eqref{eq:cdf_snr}, respectively. Again, the absolute convergence of \eqref{eq:pdf_snr}--\eqref{eq:cdf_snr} follows from observing that the Kummer confluent hypergeometric function $_1F_1$ in \eqref{eq:km:ub} is finite for any choice of parameters in our framework. Moreover, we obtain the upper bound on the truncation error as
\begin{align}
    \mathcal{E}_{\kappa}(\epsilon) < \ & \left(\frac{(\kappa +1) \mu}{\exp(\kappa)} \right)^{N \mu} \frac{8 N \Gamma (\mu +\epsilon) w^{N \mu+\epsilon-1+\zeta}}{7 \Gamma (\epsilon) \Gamma (N \mu +\epsilon+\zeta) \hat{w} ^{N \mu+\epsilon}} \nonumber \\
    & \times K^{\epsilon} \, _2F_2\bigg(1,\mu +\epsilon;\epsilon,N \mu+\epsilon+\zeta;\frac{K w}{ \hat{w}}\bigg). \label{eq:kappa:truncerror}
\end{align}

\begin{figure}[t]
\centering
\includegraphics{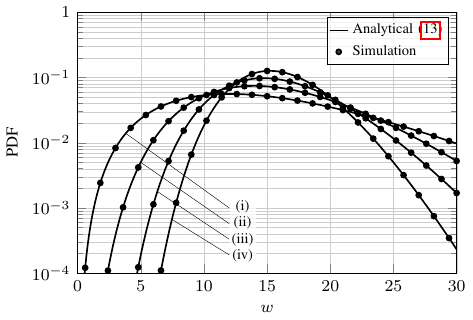}
\caption{\ac{PDF} of the sum of squared \ac{i.i.d.} \Ehm{} \acp{RV}, with $N = 16$, $\hat{w} = 1$, and the following combinations of parameters: (i) $\eta \to 0$, $\mu = 0.5$, $p = 1$ (Nakagami-$m$); (ii) $\eta = 0.6$, $\mu = 0.5$, $p = 0.5$; (iii) $\eta = 0.25$, $\mu = 1.25$, $p = 1.1$, and (iv) $\eta = 1.5$, $\mu =  2$, $p = 0.5$.}
\label{fig:pdf-hm}
\end{figure}

\subsection{Computational Complexity and Implementation}

\begin{figure}[t]
\centering
\includegraphics[]{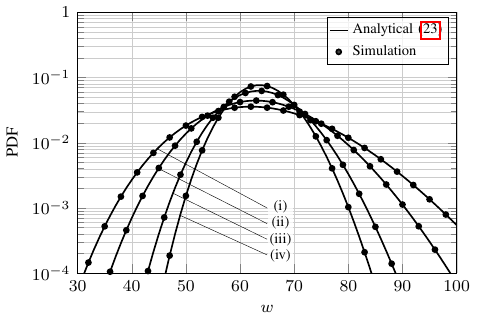}
\caption{\ac{PDF} of the sum of squared \ac{i.i.d.} \km{} \acp{RV}, with $N=64$, $\hat{w}=1$, and the following combinations of parameters: (i) $\kappa \rightarrow 0$, $\mu = 0.5$ (Nakagami-$m$); (ii) $\kappa = 1.5$, $\mu = 0.5$; (iii) $\kappa = 1.5$, $\mu = 1$ (Rice); and (iv) $\kappa = 1.5$, $\mu = 1.5$.}
\label{fig:pdf-kappa}
\end{figure}

As indicated in \eqref{eq:eta:hm} and \eqref{eq:km}, we rely on recursion to numerically evaluate the derived expressions. A naive implementation of the proposed framework would compute the same elements multiple times, greatly increasing the computational complexity when employing a large number of terms. For instance, by using the recursion tree method, the number of computations of the type of \eqref{eq:eta:hm} (resp. \eqref{eq:km}) for the \Ehm{} (resp. \km{}) distribution amounts to $2^{\epsilon} - 1$. To reduce the computational complexity, we resort to memoization, which is a code optimization method that caches the value of every new recursive element, preventing the framework from computing it multiple times. With memoization, the number of computations of the type of \eqref{eq:eta:hm} or \eqref{eq:km} reduces to $\epsilon (\epsilon + 1) - 1$, enabling fast evaluation even with a large number of terms. Figs.~\ref{fig:pdf-hm} and~\ref{fig:pdf-kappa} plot the analytical and simulated \acp{PDF} for the \Ehm{} and \km{} distributions, respectively, with $N=16$ for the former and $N=64$ for the latter, $\hat{w} =1$, and different combinations of fading parameters. The analytical \acp{PDF} are computed via \eqref{eq:eta:pdf_snr} and \eqref{eq:pdf_snr} with truncation at $\epsilon = 250$ terms, and strikingly agree with the simulations. Remarkably, the computation time for each curve in the plots peaks approximately at $0.5$~s and the maximum absolute error with respect to the exact solutions in~\cite{Mil08,Bad21} is in the order of $10^{-16}$.

\section{Performance Metrics}\label{sec:metrics}

In this section, considering the system model and the results for the \Ehm{} and \km{} distributions derived in Section~\ref{sec:newrep}, we obtain new expressions for the outage and coverage probability, \ac{BEP} for coherent binary modulations, and \ac{SEP} for \ac{$M$-PSK} and \ac{$M$-QAM}. To evaluate the system's performance at high \ac{SNR}, we further derive the asymptotic expressions for these metrics.

\subsection{Outage and Coverage Probabilities}

The outage probability is defined as the probability that the instantaneous \ac{SNR} drops below a given threshold. In this setting, the outage probability is formulated in terms of the \ac{CDF} of the instantaneous \ac{SNR}, i.e., $P_\textrm{out} = F_{\gamma}(\gamma_\textrm{th})$, where $\gamma_\textrm{th}$ is the \ac{SNR} threshold. Hence, the outage probability for the \Ehm{} and \km{} models is readily available from \eqref{eq:eta:cdf_snr} and \eqref{eq:cdf_snr}, where $\hat{w}$ is the scale parameter that can be used to model the path loss. The coverage probability is complementary to the outage probability and is thus formulated in terms of the complementary \ac{CDF} of the instantaneous \ac{SNR}, i.e., $P_\textrm{cov} = 1 - P_\textrm{out} = 1- F_{\gamma}(\gamma_\textrm{th})$.

To derive the asymptotic expressions for the outage probability at high \ac{SNR}, as done in~\cite{Alm23b}, we use only the first term of the summation (corresponding to $m=0$) in \eqref{eq:eta:cdf_snr} and \eqref{eq:cdf_snr}, resulting in
\begin{align} \label{eq:eta_Pout_as}
     P_{\textrm{out}} \sim \frac{1}{\Gamma (N \mu  +1)} \bigg(\left(\frac{p}{\eta }\right)^{\frac{p}{1+p}} \frac{\xi \gamma_\textrm{th}}{\hat{w}} \bigg)^{N\mu}
\end{align}
for the \Ehm{} model\footnote{Note that \eqref{eq:eta_Pout_as} can also be obtained as a special case of the expression provided in~\cite{Bad21} for the sum of squared independent but not identically distributed \Ehm{} \acp{RV}.} and in
\begin{align} \label{eq:kappa_Pout_as}
    P_{\textrm{out}} \sim \frac{1}{\Gamma (N \mu  +1)} \left(\frac{(\kappa +1) \mu \gamma_\textrm{th}}{\exp(\kappa) \hat{w}}\right)^{N \mu }
\end{align}
for the \km{} model.

\subsection{\ac{BEP} for Coherent Binary Modulations}

In this section, we focus on the (uncoded) \ac{BEP} for coherent binary modulations. First, we obtain new expressions for the \ac{BEP} by resorting to the \acp{PDF} in \eqref{eq:eta:pdf_snr} and \eqref{eq:pdf_snr} for the \Ehm{} and \km{} models, respectively, along with their asymptotic expressions at high \ac{SNR}. Then, by deriving an upper bound on the \ac{BEP}, we determine the convergence conditions for these expressions, which prove to be quite restrictive. To guarantee the convergence for any choice of parameters, we thus propose alternative expressions for the \ac{BEP} based on the \acp{PDF} in \eqref{eq:eta:pdf_snr2} and \eqref{eq:kappa:pdf_snr2} for the \Ehm{} and \km{} models, respectively. Lastly, we obtain an upper bound on the truncation error.

The \ac{BEP} for coherent binary modulations is defined as~\cite[Eq.~(9.3)]{Sim05}
\begin{align} \label{eq:pb}
    \textrm{BEP} = \frac{1}{2}\int_{0}^{\infty}\mathrm{erfc}(\sqrt{g_{\textrm{b}} w}) f_{W}(w) \diff w,
\end{align}
with $g_{\textrm{b}} = 1$ for coherent \ac{BPSK}, $g_{\textrm{b}} = 1/2$ for coherent orthogonal \ac{BFSK}, and $g_{\textrm{b}} = 0.715$ for coherent \ac{BFSK} with minimal correlation~\cite{Sim05}. In our case, $f_{W}(w)$ is replaced by the \acp{PDF} in \eqref{eq:eta:pdf_snr} and \eqref{eq:pdf_snr} (or \eqref{eq:eta:pdf_snr2} and \eqref{eq:kappa:pdf_snr2}).

\smallskip

\textit{\textbf{Approach 1.}} By substituting \eqref{eq:eta:pdf_snr} and \eqref{eq:pdf_snr} into \eqref{eq:pb}, we obtain the \ac{BEP} as
\begin{align}
    \textrm{BEP}_{\eta} = \ & \frac{1}{2\sqrt{\pi}}\Bigg( \frac{\big(\frac{p}{\eta }\big)^{\frac{\mu  p}{1+p}}\xi^{\mu}}{\Gamma \big(\frac{\mu  p}{1+p}\big)} \Bigg)^N \sum _{m=0}^{\infty } 
    \left(\frac{1}{g_{\textrm{b}} \hat{w}}\right)^{N \mu + m} \nonumber \\
    & \times \frac{ \Gamma \big(N \mu + m +\frac{1}{2}\big) h_m}{  \Gamma (N \mu + m +1) } \label{eq:eta:pb}
\end{align}
for the \Ehm{} model and as
\begin{align}
    \textrm{BEP}_{\kappa} = \ & \frac{1}{2\sqrt{\pi}} \left(\frac{(\kappa +1) \mu}{\exp\left(\kappa\right)}\right)^{N \mu}\sum _{m=0}^{\infty }  \left(\frac{1}{g_{\textrm{b}} \hat{w} }\right)^{N \mu + m} \nonumber \\
    & \times \frac{ \Gamma \big(N \mu + m +\frac{1}{2}\big)k_m}{\Gamma (N \mu + m +1)} \label{eq:kappa:pb}
\end{align}
for the \km{} model.

To derive the asymptotic expressions for the \ac{BEP} at high \ac{SNR}, we use only the first term of the summation (corresponding to $m=0$) in \eqref{eq:eta:pb} and \eqref{eq:kappa:pb}, resulting in
\begin{align}\label{eq:eta:pb-asymp}
    \textrm{BEP}_{\eta} \sim \ \frac{\Gamma \big(N \mu +\frac{1}{2}\big)}{2\sqrt{\pi}  \Gamma \big(N \mu +1\big)}\bigg( \bigg(\frac{p}{\eta }\bigg)^{\frac{ p}{1+p}}\frac{\xi}{g_{\textrm{b}}\hat{w}} \bigg)^{N\mu}
\end{align}
for the \Ehm{} model and 
\begin{align}\label{eq:kappa:pb-asymp}
    \textrm{BEP}_{\kappa} \sim \ \frac{ \Gamma \big(N \mu  +\frac{1}{2}\big)}{2\sqrt{\pi} \Gamma \big(N \mu  +1\big)} \left(\frac{(\kappa +1) \mu}{\exp\left(\kappa\right) g_{\textrm{b}} \hat{w}}\right)^{N \mu}
\end{align}
for the \km{} model.

Now, following similar steps as in Section~\ref{sec:trunc}, we upper bound \eqref{eq:eta:pb} and \eqref{eq:kappa:pb} as
\begin{align}
    \mathcal{C}_{\eta} < \ & 
    \frac{ |h_0| }{\hat{w} ^{N \mu}}  \Bigg(\frac{\Gamma\big(N\mu +\frac{1}{2}\big)}{\Gamma(N\mu + 1)} +\frac{8 N \Gamma (\mu +1) \Gamma \left(N \mu +\frac{3}{2}\right)}{7 \Gamma (N \mu + 2)} \nonumber \\
    & \times \frac{w \xi  p}{\hat{w} \eta } \, _2F_1 \bigg(\mu +1;N \mu +\frac{3}{2};N \mu  +2;\frac{w \xi  p}{\hat{w} \eta }\bigg)\Bigg), \label{eq:eta:pb-converg}
\end{align}
and
\begin{align}
    \mathcal{C}_{\kappa} < \ & 
    \frac{1 }{\hat{w} ^{N \mu}}  \Bigg(\frac{\Gamma\big(N\mu +\frac{1}{2}\big)}{\Gamma(N\mu + 1)} +\frac{8 N \Gamma (\mu +1) \Gamma \left(N \mu +\frac{3}{2}\right)}{7 \Gamma (N \mu + 2)} \nonumber \\
    & \times \frac{K}{g_{\textrm{b}} \hat{w} } \, _2F_1 \bigg(\mu +1;N \mu +\frac{3}{2};N \mu  +2;\frac{K}{g_{\textrm{b}}\hat{w} }\bigg)\Bigg), \label{eq:kappa:pb-converg}
\end{align}
respectively.
Despite the strict convergence interval of the Gauss hypergeometric function $_2F_1$, common mathematical packages such as Mathematica and SciPy provide solutions to arguments outside the convergence interval through its analytical continuation. In our case, this feature could not be straightforwardly exploited for the summations in \eqref{eq:eta:pb} and \eqref{eq:kappa:pb}, restricting their convergence to choice of parameters satisfying $\big| \xi p /(\eta g_{\text{b}}\hat{w})\big| < 1$ and $\big|K/ (g_{\text{b}} \hat{w})\big| < 1$, respectively. To circumvent this issue and guarantee the convergence for any choice of parameters, we propose alternative expressions in the following.

\smallskip

\begin{figure*}[t]
\setcounter{equation}{52}
\begin{align}
     \textrm{SEP}^{M\textrm{-PSK}}_{\eta} = \ & \Bigg( \frac{\big(\frac{p}{\eta }\big)^{\frac{\mu  p}{1+p}}}{\Gamma \big(\frac{\mu  p}{1+p}\big)} \Bigg)^{N  } \sum _{m=0}^{\infty} \tilde{h}_m  \Bigg(\frac{\cos \left(\frac{\pi }{M}\right)}{\pi}\left(\frac{1}{1+\frac{g_{\textrm{p}}\hat{w}}{\xi}}\right)^{N\mu+m} F_1\Bigg(\frac{1}{2};\frac{1}{2}-N \mu-m,N \mu+ m;\frac{3}{2};\cos ^2\bigg(\frac{\pi }{M}\bigg),\frac{ \cos ^2\left(\frac{\pi }{M}\right)}{1 +\frac{g_{\textrm{p}} \hat{w}}{\xi} }\Bigg) \nonumber \\
     & +\left(\frac{\xi}{g_{\textrm{p}}\hat{w}}\right)^{N\mu+m} \frac{ \Gamma \big(N \mu + m +\frac{1}{2}\big)}{2\sqrt{\pi}\Gamma (N \mu + m +1)} \, _2F_1\left(N \mu+m,N \mu+m+\frac{1}{2};N \mu+m+1;-\frac{\xi}{g_{\textrm{p}}\hat{w}}\right)\Bigg) \label{eq:eta:psk-2}
\end{align}
\vspace{-1mm}
\hrulefill
\vspace{-2mm}
\end{figure*}
%
\begin{figure*}[t]
\setcounter{equation}{54}
\begin{align}
     \textrm{SEP}^{M\textrm{-PSK}}_{\kappa} = \ & \exp(-N\kappa\mu) \sum _{m=0}^{\infty} \tilde{k}_m  \Bigg(\frac{\cos \left(\frac{\pi }{M}\right)}{\pi}\left(\frac{1}{1+\frac{g_{\textrm{p}}\hat{w}}{\tilde{K}}}\right)^{N\mu+m} F_1\Bigg(\frac{1}{2};\frac{1}{2}-N \mu-m,N \mu+ m;\frac{3}{2};\cos ^2\bigg(\frac{\pi }{M}\bigg),\frac{ \cos ^2\left(\frac{\pi }{M}\right)}{1 +\frac{g_{\textrm{p}} \hat{w}}{\tilde{K}} }\Bigg) \nonumber \\
     & +\bigg(\frac{\tilde{K}}{g_{\textrm{p}}\hat{w}}\bigg)^{N\mu+m} \frac{ \Gamma \big(N \mu + m +\frac{1}{2}\big)}{2\sqrt{\pi}\Gamma (N \mu + m +1)} \, _2F_1\bigg(N \mu+m,N \mu+m+\frac{1}{2};N \mu+m+1;-\frac{\tilde{K}}{g_{\textrm{p}}\hat{w}}\bigg)\Bigg) \label{eq:kappa:psk-2}
\end{align}
\setcounter{equation}{46}
\vspace{-1mm}
\hrulefill
\vspace{-2mm}
\end{figure*}

\textit{\textbf{Approach 2.}} By substituting \eqref{eq:eta:pdf_snr2} and \eqref{eq:kappa:pdf_snr2} into \eqref{eq:pb}, we obtain alternative \ac{BEP} expression as
\begin{align}
    \textrm{BEP}_{\eta} = \ & \Bigg( \frac{\big(\frac{p}{\eta }\big)^{\frac{\mu  p}{1+p}}}{\Gamma \big(\frac{\mu  p}{1+p}\big)} \Bigg)^N  \sum _{m=0}^{\infty } \frac{ \Gamma \big(N \mu \! + \! m \! + \! \frac{1}{2}\big) \big(\frac{\xi}{g_{\textrm{b}} \hat{w}}\big)^{N \mu + m} \tilde{h}_m}{ 2\sqrt{\pi} \Gamma (N \mu + m + 1)} \nonumber\\
    &\times \, _2F_1 \bigg(N \mu \! + \! m,N \mu \!+\! m\!+\!\frac{1}{2};N \mu \! + \!m+1;\frac{ -\xi }{g_{\textrm{b}} \hat{w} }\bigg) \label{eq:eta:pbfinal2}
\end{align}
for the \Ehm{} distribution (cf. \eqref{eq:eta:pb}) and as
\begin{align}
    \textrm{BEP}_{\kappa} = \ & \sum _{m=0}^{\infty } \frac{ \Gamma \big(N \mu \! + \! m \! + \!\frac{1}{2}\big) \big(\frac{\tilde{K}}{g_{\textrm{b}} \hat{w}}\big)^{N \mu + m} \tilde{k}_m}{ 2\sqrt{\pi} \exp(N\kappa\mu) \Gamma (N \mu \!+ \!m \! + \! 1)} \, _2F_1\bigg(N \mu \! + \! m, \nonumber \\
    & N \mu + m+\frac{1}{2};N \mu + m+1;\frac{ -\tilde{K} }{g_{\textrm{b}}  \hat{w} }\bigg) \label{eq:kappa:pbfinal2}
\end{align}
for the \km{} distribution as (cf. \eqref{eq:kappa:pb}). As commented above, \eqref{eq:eta:pbfinal2} and \eqref{eq:kappa:pbfinal2} allow to evaluate the \ac{BEP} for any choice of parameters at the cost of a slight increase in mathematical and computational complexity due to the presence of the Gauss hypergeometric function $_2F_1$. However, the latter can be drastically reduced by using its recurrence property~\cite[Eq. (15.5.E19)]{NIST:DLMF}.

\smallskip

\textit{\textbf{Upper Bound on the Truncation Error.}} An upper bound on the truncation error for the \ac{BEP} is obtained by replicating the steps in Section~\ref{sec:trunc}. The following expressions are suitable for both the approaches described above. Hence, an upper bound on the truncation error for the \ac{BEP} is given by 
\begin{align}
    \mathcal{E}_{\eta}(\epsilon) < \ & \, _3F_2\left(N \mu +\epsilon+\frac{1}{2},\mu +\epsilon,1;N \mu +\epsilon+1,\epsilon; \frac{ \xi  p}{g_{\textrm{b}}\hat{w} \eta  }\right) \nonumber \\
    & \times  \frac{4 N\Gamma (\mu +\epsilon) \Gamma \big(N \mu +\epsilon+\frac{1}{2}\big)}{7\sqrt{\pi}\Gamma \left(\epsilon\right) \Gamma(N \mu +\epsilon+1)} \bigg(\frac{ \xi  p}{ \eta g_{\textrm{b}}\hat{w}   }\bigg)^{\epsilon} \nonumber \\
    & \times \Bigg(\frac{\big(\frac{p}{\eta }\big)^{\frac{\mu  p}{1+p}}}{\Gamma\big(\frac{\mu p}{1+p}\big) \left(g_{\textrm{b}}\hat{w} \right)^{\mu}}\Bigg)^N \label{eq:eta:bep-trunc}
\end{align}
for the \Ehm{} model and 
\begin{align}
   \mathcal{E}_{\kappa} (\epsilon) < \ & \, _3F_2\left(N \mu+\epsilon+\frac{1}{2},\mu +\epsilon,1;N \mu+\epsilon+1,\epsilon;\frac{K}{g_{\textrm{b}} \hat{w} }\right) \nonumber \\
    & \times \frac{4 N\Gamma \left(\mu +\epsilon\right) \Gamma \big(N \mu +\epsilon+\frac{1}{2}\big)}{7 \sqrt{\pi} \Gamma \left(\epsilon\right) \Gamma \left(N \mu +\epsilon+1\right)} \bigg(\frac{K}{g_{\textrm{b}} \hat{w}}\bigg)^{\epsilon} \nonumber \\
    &\times \left(\frac{(\kappa +1) \mu }{\exp (\kappa ) g_{\textrm{b}} \hat{w}}\right)^{N \mu} \label{eq:kappa:bep-trunc}
\end{align}
for the \km{} model.

\subsection{\ac{SEP} for \ac{$M$-PSK} and \ac{$M$-QAM}}

In this section, we focus on the (uncoded) \ac{SEP} for \ac{$M$-PSK} and \ac{$M$-QAM}. Specifically, we obtain new expressions for the \ac{SEP} by resorting to the \acp{MGF} in \eqref{eq:eta:mgf_sum} and \eqref{eq:eta:mgf-pb2} for the \Ehm{} model, and in \eqref{eq:kappa:mgf} and \eqref{eq:km:mgf-pb2} for the \km{} model. As for the \ac{BEP} in the previous section, the first expression for each model has lower computational complexity but quite restrictive convergence conditions (approach~1), whereas the second has slightly higher computational complexity and no restrictions on the convergence (approach~2). From the first set of expressions, we also derive the asymptotic \ac{SEP} at high \ac{SNR}.

\begin{figure*}[t]
\setcounter{equation}{59}
\begin{align}
     \textrm{SEP}^{M\textrm{-QAM}}_{\eta} = \ & \Bigg( \frac{\left(\frac{p}{\eta }\right)^{\frac{\mu  p}{1+p}}}{\Gamma \big(\frac{\mu  p}{1+p}\big)} \Bigg)^{N  } \sum _{m=0}^{\infty} \frac{2 (\sqrt{M}-1) \tilde{h}_m}{M \pi } \Bigg( \frac{\sqrt{\pi } \Gamma (N \mu + m +\frac{1}{2})}{\Gamma (N \mu + m )}\bigg(\frac{\sqrt{M}}{ (-1)^{N\mu + m}} B_{\frac{-\xi }{g_{\textrm{q}} \hat{w} }}\bigg(N\mu +m, \frac{1}{2}-N\mu-m\bigg) \nonumber \\
     &-(\sqrt{M}-1) B\bigg(\frac{\xi}{g_{\textrm{q}} \hat{w}+\xi};N \mu + m ,\frac{1}{2}\bigg)\bigg)+ \sqrt{2} (\sqrt{M}-1)\left(\frac{\xi}{g_{\textrm{q}} \hat{w}+\xi}\right)^{N\mu +m} \nonumber \\
     &\times F_1\left(\frac{1}{2};\frac{1}{2}-N\mu-m,N\mu + m;\frac{3}{2};\frac{1}{2},\frac{\xi}{2 \left(g_{\textrm{q}} \hat{w} +\xi\right)}\right)\Bigg) \label{eq:eta:qam-2}
\end{align}
\vspace{-1mm}
\hrulefill
\vspace{-1mm}
\end{figure*}
%
\begin{figure*}[t]
\setcounter{equation}{61}
\begin{align}
     \textrm{SEP}^{M\textrm{-QAM}}_{\kappa} = \ &  \sum _{m=0}^{\infty} \frac{2 (\sqrt{M}-1) \tilde{k}_m}{M \pi \exp(N\kappa\mu) } \Bigg( \frac{\sqrt{\pi } \Gamma (N \mu + m +\frac{1}{2})}{\Gamma (N \mu + m )}\bigg(\frac{\sqrt{M}}{ (-1)^{N\mu + m}} B\bigg({\frac{-\tilde{K} }{ g_{\textrm{q}} \hat{w}} };N\mu +m, \frac{1}{2}-N\mu-m\bigg) \nonumber \\
     &-(\sqrt{M}-1) B\bigg({\frac{\tilde{K} }{\tilde{K} +  g_{\textrm{q}} \hat{w}}};N \mu + m ,\frac{1}{2}\bigg)\bigg)+ \sqrt{2} (\sqrt{M}-1)\bigg(\frac{\tilde{K}}{
     \tilde{K} + g_{\textrm{q}} \hat{w}}\bigg)^{N\mu +m} \nonumber \\
     &\times F_1\bigg(\frac{1}{2};\frac{1}{2}-N\mu-m,N\mu + m;\frac{3}{2};\frac{1}{2},\frac{\tilde{K}}{2 ( \tilde{K} + g_{\textrm{q}} \hat{w})}\bigg)\Bigg) \label{eq:kappa:qam-2}
\end{align}
\setcounter{equation}{50}
\vspace{-1mm}
\hrulefill
\vspace{-1mm}
\end{figure*}

\smallskip

\textit{\textbf{\boldmath\ac{$M$-PSK} Modulation.}} The \ac{SEP} for \ac{$M$-PSK} is defined as~\cite[Eq.~(9.15)]{Sim05}
\begin{align}\label{eq:psk}
    \textrm{SEP}^{M\textrm{-PSK}} = \frac{1}{\pi} \int_{0}^{\frac{(M-1)}{M}\pi} \mathcal{M}_{W} \left(\frac{g_{\textrm{p}}}{\sin^2 \phi}\right) \diff\phi, 
\end{align}
with $g_{\textrm{p}} = \sin^{2} \left(\pi/M\right)$.

For the \Ehm{} model, the \ac{SEP} can be expressed as
\begin{align}
    \textrm{SEP}^{M\textrm{-PSK}}_{\eta} = \, & \Bigg( \frac{\big(\frac{p}{\eta }\big)^{\frac{\mu  p}{1+p}}\xi^{\mu}}{\Gamma \big(\frac{\mu  p}{1+p}\big)} \Bigg)^{N  } \sum _{m=0}^{\infty } \left(\frac{1}{g_{\textrm{p}} \hat{w}}\right)^{N \mu + m} \frac{h_m}{2\pi} \nonumber \\
    & \times \Bigg(\frac{2 \sqrt{\pi } \Gamma \big(N \mu + m +\frac{1}{2}\big)}{\Gamma (N \mu + m +1)} \nonumber \\
    & -B\left(\sin ^2\bigg(\frac{\pi }{M}\bigg); N \mu + m +\frac{1}{2},\frac{1}{2}\right)\Bigg) \label{eq:eta:psk-1}
\end{align}
for the \ac{MGF} in \eqref{eq:eta:mgf_sum} and as in \eqref{eq:eta:psk-2} at the top of the page for the \ac{MGF} in \eqref{eq:eta:mgf-pb2}.

For the \km{} model, the \ac{SEP} can be expressed as
\setcounter{equation}{53}
\begin{align}
    \textrm{SEP}^{M\textrm{-PSK}}_{\kappa} = \ & \left(\frac{(\kappa +1) \mu }{\exp(\kappa) }\right)^{N \mu }\sum _{m=0}^{\infty } \left(\frac{1}{g_{\textrm{p}} \hat{w}}\right)^{N \mu + m} \frac{k_m}{2\pi} \nonumber \\
    & \times \Bigg(\frac{2 \sqrt{\pi } \Gamma \big(N \mu + m +\frac{1}{2}\big)}{\Gamma (N \mu + m +1)} \nonumber \\
    & -B\left(\sin ^2\bigg(\frac{\pi }{M}\bigg); N \mu + m +\frac{1}{2},\frac{1}{2}\right)\Bigg) \label{eq:kappa:psk-1}
\end{align}
for the \ac{MGF} in \eqref{eq:kappa:mgf} and as in \eqref{eq:kappa:psk-2} at the top of the page for the \ac{MGF} in \eqref{eq:km:mgf-pb2}.

To derive the asymptotic expressions for the \ac{SEP} at high \ac{SNR}, we use only the first term of the summation (corresponding to $m=0$) in \eqref{eq:eta:psk-1} and \eqref{eq:kappa:psk-1}, resulting in
\setcounter{equation}{55}
\begin{align}
    \textrm{SEP}^{M\textrm{-PSK}}_{\eta} \sim \ & \frac{1}{2\pi}\bigg( \left(\frac{p}{\eta }\right)^{\frac{p}{1+p}} \frac{\xi}{g_{\textrm{p}} \hat{w}} \bigg)^{N\mu} \Bigg(\frac{2 \sqrt{\pi } \Gamma \big(N \mu +\frac{1}{2}\big)}{\Gamma (N \mu +1)} \nonumber \\
    & -B\left(\sin ^2\bigg(\frac{\pi }{M}\bigg); N \mu +\frac{1}{2},\frac{1}{2}\right)\Bigg) \label{eq:eta:psk-asymp}
\end{align}
for the \Ehm{} model and
\begin{align}
\textrm{SEP}^{M\textrm{-PSK}}_{\kappa} \sim \ & \frac{1}{2\pi}\left(\frac{(\kappa +1) \mu }{\exp(\kappa) g_{\textrm{p}}\hat{w} }\right)^{N \mu } \Bigg(\frac{2 \sqrt{\pi } \Gamma \big(N \mu +\frac{1}{2}\big)}{\Gamma (N \mu +1)} \nonumber \\
    & -B\left(\sin ^2\bigg(\frac{\pi }{M}\bigg); N \mu +\frac{1}{2},\frac{1}{2}\right)\Bigg) \label{eq:kappa:psk-asymp}
\end{align}
for the \km{} model.

\smallskip

\textit{\textbf{\boldmath\ac{$M$-QAM} Modulation.}} The \ac{SEP} for \ac{$M$-QAM} is defined as~\cite[Eq.~(9.21)]{Sim05}
\begin{align}
    \textrm{SEP}^{M\textrm{-QAM}} = \ & \frac{4}{\pi}\left(1-\frac{1}{\sqrt{M}}\right) \int_{0}^{\pi/2}  \mathcal{M}_{W} \left(\frac{g_{\textrm{q}}}{\sin^2 \phi}\right) \diff\phi \nonumber \\
    & - \frac{4}{\pi}\left(1-\frac{1}{\sqrt{M}}\right)^2 \int_{0}^{\pi/4} \mathcal{M}_{W} \left(\frac{g_{\textrm{q}}}{\sin^2 \phi}\right) \diff\phi, \label{eq:qam}
\end{align}
with $g_{\textrm{q}} = 3/\big(2 (M-1)\big)$.

For the \Ehm{} model, the \ac{SEP} can be expressed as
\begin{align}
    \textrm{SEP}^{M\textrm{-QAM}}_{\eta} = \ & \frac{2}{\sqrt{\pi}}\Bigg( \frac{\big(\frac{p}{\eta }\big)^{\frac{\mu  p}{1+p}} \xi^{\mu}}{\Gamma \big(\frac{\mu  p}{1+p}\big)} \Bigg)^{N  } \sum _{m=0}^{\infty } \left(\frac{1}{g_{\textrm{q}}\hat{w}}\right)^{N\mu + m} \nonumber \\ 
    & \times h_m \left(1-\frac{1}{\sqrt{M}}\right)\Bigg( \frac{ \Gamma \big(N \mu + m +\frac{1}{2}\big)}{\Gamma (N \mu + m +1)} \nonumber \\
    & - \frac{1-\frac{1}{\sqrt{M}}}{\sqrt{\pi} } B\bigg(\frac{1}{2};N \mu + m+\frac{1}{2},\frac{1}{2}\bigg)\Bigg) \label{eq:eta:qam-1}
\end{align}
for the \ac{MGF} in \eqref{eq:eta:mgf_sum} and as in \eqref{eq:eta:qam-2} at the top of the next page for the \ac{MGF}s in \eqref{eq:eta:mgf-pb2}.

For the \km{} model, the \ac{SEP} can be expressed as
\setcounter{equation}{60}
\begin{align}
    \textrm{SEP}^{M\textrm{-QAM}}_{\kappa} = \ & \frac{2}{\sqrt{\pi}}\left(\frac{(\kappa +1) \mu }{\exp(\kappa) }\right)^{N \mu } \sum_{m=0}^{\infty} \left(\frac{1}{g_{\textrm{q}}\hat{w}}\right)^{N\mu+m} \nonumber \\ 
    & \times k_{m} \left(1-\frac{1}{\sqrt{M}}\right)\Bigg(\frac{ \Gamma \big(N \mu + m +\frac{1}{2}\big)}{\Gamma (N \mu + m +1)} \Bigg. \nonumber \\
    & - \frac{1-\frac{1}{\sqrt{M}}}{\sqrt{\pi} } B\bigg(\frac{1}{2};N \mu + m+\frac{1}{2},\frac{1}{2}\bigg)\Bigg) \label{eq:kappa:qam-1}
\end{align}
for the \ac{MGF} in \eqref{eq:kappa:mgf} and as in \eqref{eq:kappa:qam-2} at the top of the next page for the \ac{MGF} in \eqref{eq:km:mgf-pb2}.

To derive the asymptotic expressions for the \ac{SEP} at high \ac{SNR}, we use only the first term of the summation (corresponding to $m=0$) in \eqref{eq:eta:qam-1} and \eqref{eq:kappa:qam-1}, resulting in
\setcounter{equation}{62}
\begin{align}
    \textrm{SEP}^{M\textrm{-QAM}}_{\eta} \sim \ & \frac{2 \big(1-\frac{1}{\sqrt{M}}\big)}{\sqrt{\pi}}\bigg( \left(\frac{p}{\eta }\right)^{\frac{  p}{1+p}} \frac{\xi}{g_{\textrm{q}}\hat{w}} \bigg)^{N\mu} \Bigg( \frac{ \Gamma \big(N \mu +\frac{1}{2}\big)}{\Gamma (N \mu +1)} \nonumber \\
    & - \frac{1-\frac{1}{\sqrt{M}}}{\sqrt{\pi} } B\bigg(\frac{1}{2};N \mu + \frac{1}{2},\frac{1}{2}\bigg)\Bigg) \label{eq:eta:qam-asymp}
\end{align}
for the \Ehm{} model and
\begin{align}
    \textrm{SEP}^{M\textrm{-QAM}}_{\kappa} \sim \ & \frac{2 \big(1-\frac{1}{\sqrt{M}}\big)}{\sqrt{\pi}}\left(\frac{(\kappa +1) \mu }{\exp(\kappa) g_{\textrm{q}}\hat{w} }\right)^{N \mu } \Bigg(\frac{ \Gamma \big(N \mu +\frac{1}{2}\big)}{\Gamma (N \mu +1)} \Bigg. \nonumber \\
    & - \frac{1-\frac{1}{\sqrt{M}}}{\sqrt{\pi} } B\bigg(\frac{1}{2};N \mu +\frac{1}{2},\frac{1}{2}\bigg)\Bigg) \label{eq:kappa:qam-asymp}
\end{align}
for the \km{} model.

\begin{table}[t!]
\centering
\footnotesize
\SetTblrInner{rowsep=1pt}
\begin{tblr}{|l|Q[c,8mm]|Q[c,8mm]|Q[c,8mm]|Q[c,8mm]|}
    \hline
                                                & \SetCell[c=2]{c}{\textbf{Extended} {\boldmath \hm}} &     & \SetCell[c=2]{c}{{\boldmath \km}} & \\
    \hline
    \hline
    \ac{MGF}                                    & \eqref{eq:eta:mgf_sum}        & \eqref{eq:eta:mgf-pb2}    & \eqref{eq:kappa:mgf}          & \eqref{eq:km:mgf-pb2} \\
    \hline
    \ac{PDF}                                    & \eqref{eq:eta:pdf_snr}        &    \eqref{eq:eta:pdf_snr2}                       & \eqref{eq:pdf_snr}            & \eqref{eq:kappa:pdf_snr2} \\
    \hline
    \ac{CDF} (outage probability)               & \eqref{eq:eta:cdf_snr}        &   \eqref{eq:eta:cdf_snr2}                        & \eqref{eq:cdf_snr}            & \eqref{eq:kappa:cdf_snr2}\\
    \rotatebox[origin=c]{180}{$\Lsh$} high \ac{SNR}  & \eqref{eq:eta_Pout_as}        &                           & \eqref{eq:kappa_Pout_as}      & \\
    \hline
    \ac{BEP}                                    & \eqref{eq:eta:pb}             & \eqref{eq:eta:pbfinal2}   & \eqref{eq:kappa:pb}           & \eqref{eq:kappa:pbfinal2} \\
    \rotatebox[origin=c]{180}{$\Lsh$} high \ac{SNR}  & \eqref{eq:eta:pb-asymp}       &                           & \eqref{eq:kappa:pb-asymp}     & \\
    \hline
    \ac{SEP} for \ac{$M$-PSK}                   & \eqref{eq:eta:psk-1}          & \eqref{eq:eta:psk-2}      & \eqref{eq:kappa:psk-1}        & \eqref{eq:kappa:psk-2} \\
    \rotatebox[origin=c]{180}{$\Lsh$} high \ac{SNR}  & \eqref{eq:eta:psk-asymp}      &                           & \eqref{eq:kappa:psk-asymp}    & \\
    \hline
    \ac{SEP} for \ac{$M$-QAM}                   & \eqref{eq:eta:qam-1}          & \eqref{eq:eta:qam-2}      & \eqref{eq:kappa:qam-1}        & \eqref{eq:kappa:qam-2} \\
    \rotatebox[origin=c]{180}{$\Lsh$} high \ac{SNR}  & \eqref{eq:eta:qam-asymp}      &                           & \eqref{eq:kappa:qam-asymp}    & \\
    \hline
\end{tblr}
\caption{Summary of derived expressions. For each model, the expressions in the first column have lower computational complexity but quite restrictive convergence conditions, whereas those in the second column have slightly higher computational complexity and no restrictions on the convergence.}
\label{tab:expressions}
\end{table}

\section{Performance Evaluation} \label{sec:num}

In this section, we build on the proposed framework for the \Ehm{} and \km{} models to analyze the performance of a downlink system operating in FR3 and the sub-THz band, respectively. As in the system model described in Section~\ref{sec:newrep}, we consider a single-antenna user served by a multi-antenna \ac{BS} employing \ac{MRT}, with perfect and imperfect \ac{CSI} at the latter. We compare the analytical results (obtained using Mathematica on a standard off-the-shelf laptop computer) with Monte Carlo simulations. In all the plots, the lines indicate the analytical results obtained with the proposed framework (summarized in Table~\ref{tab:expressions}) and the markers represent the simulation results, which are in agreement with the analytical results.

We assume a log-distance path loss model according to which we have $\hat{w} = \frac{P_{\textrm{t}}}{\sigma^{2}} \varphi d^{-\beta}$, where $\varphi$ represents the frequency-dependent path loss factor referenced at 1~m, $d$ denotes the distance between the BS and the user, and $\beta$ indicates the path loss exponent. The path loss factor is defined as $\varphi = \big( c/(4\pi f_\textrm{c}) \big)^{2}$, where $c$ represents the speed of light and $f_\textrm{c}$ denotes the carrier frequency. The noise power is calculated as $\sigma^2 = -174 + 10\log_{10} \Omega+ \nu$ (measured in dBm), where $\Omega$ denotes the transmission bandwidth and $\nu = 5$~dB is the noise figure of the user. Unless otherwise stated, we fix $P_{\textrm{t}} = 30$~dBm and assume perfect \ac{CSI} at the \ac{BS} (i.e., $\alpha = 0$).

\subsection{FR3 System Modeled with \Ehm{} Fading} \label{sec:num_mmwave}

In this section, we evaluate the performance of an FR3 system with dominant \ac{NLoS} components using the \Ehm{} model. The \ac{NLoS}-dominated propagation is imposed through the path loss exponent $\beta = 3$. We assume that the bandwidth is $3 \%$ of the carrier frequency, i.e., $\Omega = 0.03 f_{\textrm{c}}$, as in~\cite{Kan24}. Unless otherwise stated, we assume $f_{\textrm{c}} = 15$~GHz, $d$ = 250~m, and severe fading conditions characterized by $\eta = 1.5$, $\mu = 0.5$, and $p = 0.75$. For the number of antennas, we set $N \in \{16,32,64,128,256\}$.

\begin{figure}[t]
\centering
\includegraphics[]{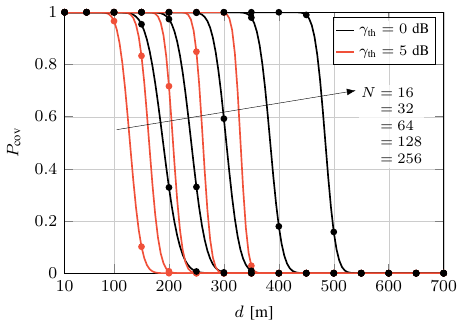}
\caption{\Ehm{} model: Downlink coverage probability versus distance, with $\eta = 1.5$, $\mu = 0.5$, $p=0.75$, $f_\textrm{c} = 15$~GHz, $P_{\textrm{t}} = 30$~dBm, $N \in \{16, 32, 64, 128, 256\}$, $\alpha=0$, and $\gamma_{\text{th}} \in \{0,5\}$~dB.}
\label{fig:hm-coverage1}
\end{figure}

Fig.~\ref{fig:hm-coverage1} depicts the coverage probability versus the distance between the \ac{BS} and the user, with \ac{SNR} threshold $\gamma_\textrm{th} \in \{0, 5\}$~dB. As expected, the coverage improves significantly as the number of antennas increases. For instance, using $N=256$ allows to achieve perfect coverage up to approximately $300$~m and $450$~m with $\gamma_\textrm{th} = 5$~dB and $\gamma_\textrm{th} = 0$~dB, respectively. The analytical results are computed with $100$ terms and the entire plot takes about $1$~s to generate. For comparison, the expression in~\cite{Bad21} takes up to $100$~s per point with $N=16$ and, as the number of antenna increases, the computation time grows significantly and eventually the expression fails to converge to the correct value.

\begin{figure}[t]
    \centering
    \includegraphics[]{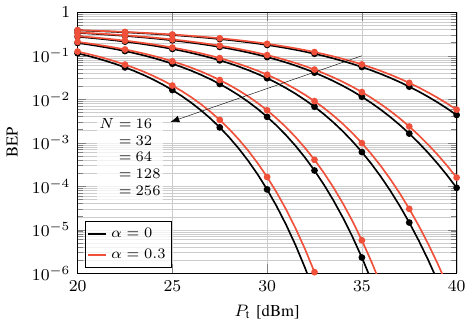}
    \caption{\Ehm{} model: Downlink \ac{BEP} for coherent \ac{BPSK} versus transmit power, with $\eta = 1.5$, $\mu = 0.5$, $p = 0.75$, $f_ \textrm{c} = 15$~GHz, $d = 250$~m, $N \in \{16, 32, 64, 128, 256\}$, and $\alpha \in \{0,0.3\}$.}
    \label{fig:eta-bep1}
\end{figure}

Fig.~\ref{fig:eta-bep1} shows the \ac{BEP} for coherent \ac{BPSK} versus the transmit power. We assume perfect and imperfect \ac{CSI} with $\alpha \in \{0,0.3\}$. Again, we emphasize the benefits of increasing the number of antennas. For $P_{\textrm{t}} = 30$~dBm and $\alpha=0$, doubling the number of antennas from $N=128$ to $N=256$ yields a $50 \times$ reduction in the \ac{BEP}, i.e., from $4\times 10^{-3}$ to $8\times 10^{-5}$. The analytical results are computed with $200$ terms, remarkably taking less than $2$~s to generate the whole plot.

\begin{figure}[t]
    \centering
    \includegraphics[]{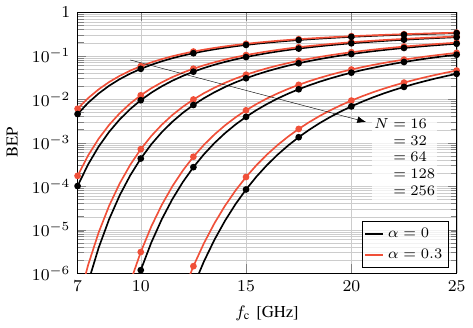}
    \caption{\Ehm{} model: Downlink \ac{BEP} for coherent \ac{BPSK} versus carrier frequency, with $\eta = 1.5$, $\mu = 0.5$, $p = 0.75$, $d = 250$~m, $P_{\textrm{t}} = 30$~dBm, $N \in \{16, 32, 64, 128, 256\}$, and $\alpha\in\{0,0.3\}$.}
    \label{fig:eta-bep2}
\end{figure}

Fig.~\ref{fig:eta-bep2} plots the \ac{BEP} for coherent \ac{BPSK} versus the carrier frequency, again with $\alpha\in \{0,0.3\}$. In this setting, the carrier frequency can be increased as the number of antennas grows. For example, fixing a target \ac{BEP} of $10^{-3}$, increasing the number of antennas from $N=32$ to $N=256$ allows to extend the carrier frequency approximately from $8$~GHz to $17$~GHz and the corresponding bandwidth from about $240$~MHz to $510$~MHz.

\begin{figure}[t]
    \centering
    \includegraphics[]{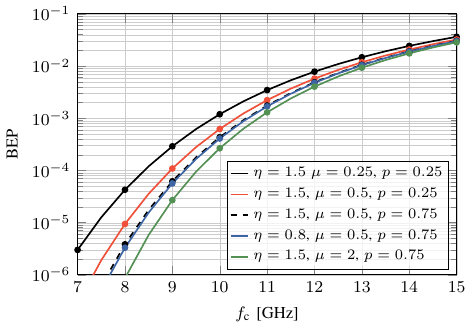}
    \caption{\Ehm{} model: Downlink \ac{BEP} for coherent \ac{BPSK} versus carrier frequency, with $\eta \in \{0.8,1.5\}$, $\mu \in \{0.25,0.5,2\}$, $p \in \{0.25,0.75\}$, $d = 250$~m, $P_{\textrm{t}} = 30$~dBm, $N= 64$, and $\alpha=0$.}
    \label{fig:eta-bep4}
\end{figure}

Fig.~\ref{fig:eta-bep4} illustrates the \ac{BEP} for coherent \ac{BPSK} versus the carrier frequency, with $N=64$. We consider five combinations of fading parameters modeling fading conditions ranging from very severe to favorable: (i) $\eta = 1.5$, $\mu = 0.25$, $p=0.25$ (black line); (ii) $\eta = 1.5$, $\mu = 0.5$, $p=0.25$ (red line); (iii) $\eta = 1.5$, $\mu = 0.5$, $p=0.75$ (black dashed line); (iv) $\eta = 0.8$, $\mu = 0.5$, $p=0.75$ (blue line); and (v) $\eta = 1.5$, $\mu = 2$, $p=0.75$ (green line). In (i), the \ac{BEP} is primarily impaired due to the small values of $\mu$ (i.e., the number of multipath clusters) and $p$ (i.e., the ratio between the numbers of multipath clusters of the in-phase and of the quadrature components). In (ii), by doubling $\mu$, the \ac{BEP} improves significantly since the array gain highly depends on $\mu$. Furthermore, increasing $p$ in (iii) yields a smaller but noteworthy improvement. In (iv), by decreasing $\eta$ (i.e., the power ratio between the in-phase and quadrature signals) such that $\eta\to p$, the \ac{BEP} decreases only marginally. Lastly, in (v), by increasing $\mu$ (again with $\eta = 1.5$), we observe a further improvement in the \ac{BEP}.

\begin{figure}[t]
    \centering
    \includegraphics[]{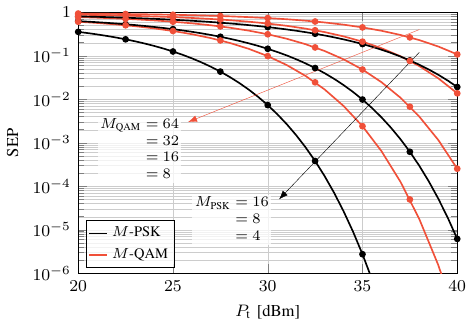}
    \caption{\Ehm{} model: Downlink \ac{SEP} for \ac{$M$-PSK} and \ac{$M$-QAM} versus transmit power, with $\eta = 1.5$, $\mu = 0.5$, $p=0.75$, $f_{\textrm{c}} = 15$~GHz, $d = 250$~m, $P_{\textrm{t}} = 30$~dBm, $N = 256$, $\alpha=0$, $M_{\textrm{PSK}} \in \{4,8,16 \}$, and $M_{\textrm{QAM}} \in \{8,16,32,64\}$.}
    \label{fig:eta-sep1}
\end{figure}

\begin{figure}[t]
    \centering
    \includegraphics[]{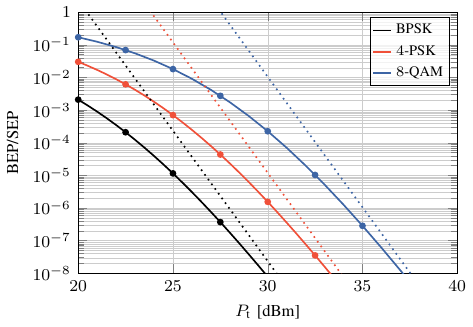}
    \caption{\Ehm{} model: Downlink BEP for coherent \ac{BPSK} and \ac{SEP} for $4$-PSK and $8$-QAM versus transmit power, with $\eta = 1.5$, $\mu = 0.5$, $p=0.75$, $f_{\textrm{c}} = 15$~GHz, $d = 50$~m, $P_{\textrm{t}} = 30$~dBm, $N = 16$, and $\alpha=0$.}
    \label{fig:eta-asympt}
\end{figure}

Fig.~\ref{fig:eta-sep1} depicts the \ac{SEP} for \ac{$M$-PSK} and \ac{$M$-QAM} versus the transmit power, with $N=256$ and modulation orders $M_{\textrm{PSK}}\in\{4,8,16\}$ and $M_{\textrm{QAM}}\in\{8,16,32,64\}$. Note that $4$-QAM is omitted since it coincides with $4$-PSK. Evidently, the \ac{SEP} worsens as the modulation order increases. Furthermore, for fixed modulation order, \ac{$M$-QAM} yields a better performance compared with \ac{$M$-PSK} due to the larger distance between the constellation points.

Lastly, Fig.~\ref{fig:eta-asympt} shows the asymptotic \ac{BEP} for coherent \ac{BPSK} along with the asymptotic \ac{SEP} for $4$-PSK and $8$-QAM at high \ac{SNR} versus the transmit power. To observe the asymptotic behavior (dotted lines), we reduce the distance and the number of antennas to $d=50$ and $N=16$, respectively. In particular, the slope is mainly dictated by the number of antennas and number of multipath clusters.

\subsection{Sub-THz Systems Modeled with \km{} Fading} \label{sec:num_subthz}

In this section, we evaluate the performance of a sub-THz system with dominant \ac{LoS} components using the \km{} model. The \ac{LoS}-dominated propagation is imposed through the path loss exponent $\beta = 2$. We assume that the bandwidth is $1 \%$ of the carrier frequency, i.e., $\Omega = 0.01 f_{\textrm{c}}$, as in~\cite{Xin22}. Unless otherwise stated, we assume $f_{\textrm{c}} = 140$~GHz, $d=300$~m, and severe fading conditions characterized by $\kappa = 0.5$ and $\mu = 0.5$. For the number of antennas, we set $N \in \{64,128,256,512,1024\}$.

\begin{figure}[t]
\centering
\includegraphics[]{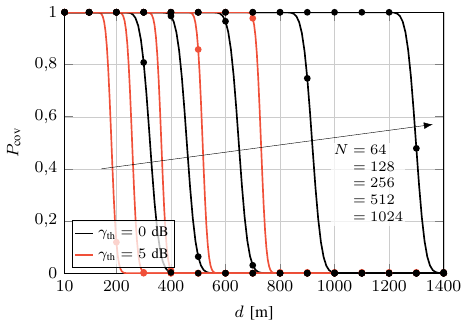}
\caption{\km{} model: Downlink coverage probability versus distance, with $\kappa = 0.5$, $\mu = 0.5$, $f_\textrm{c} = 140$~GHz, $P_{\textrm{t}} = 30$~dBm,  $N \in \{64,128,256,512,1024\}$, $\alpha=0$, and $\gamma_{\textrm{th}} \in \{0,5\}$~dB.}
\label{fig:kappa-coverage2}
\end{figure}

\begin{figure}[t]
    \centering
    \includegraphics[]{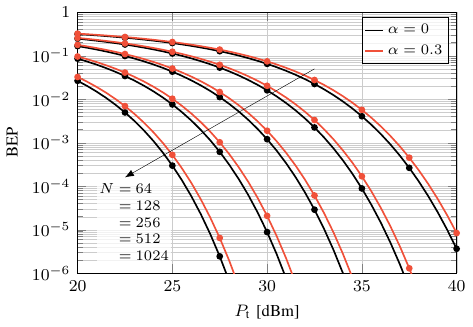}
    \caption{\km{} model: Downlink \ac{BEP} for coherent \ac{BPSK} versus transmit power, with $\kappa = 0.5$, $\mu = 0.5$, $f_\textrm{c} = 140$~GHz, $d = 300$~m, $N \in \{64, 128, 256, 512, 1024\}$, and $\alpha\in\{0,0.3\}$.}
    \label{fig:kappa-bep-pt}
\end{figure}

Fig.~\ref{fig:kappa-coverage2} plots the coverage probability versus the distance between the \ac{BS} and the user, with $\gamma_\textrm{th} \in \{0, 5\}$~dB. 
Adopting $N=1024$ allows to compensate for the strong path loss at $140$~GHz and achieve perfect coverage up to approximately $700$~m and $1200$~m for $\gamma_{\textrm{th}} =5$~dB and $\gamma_{\textrm{th}} =0$~dB, respectively. Note that this extensive coverage is primarily enabled by the truly massive number of antennas combined with the \ac{LoS}-dominated propagation, in contrast with the \ac{NLoS}-dominated propagation characterizing the FR3 system in Section~\ref{sec:num_mmwave} (cf. Fig.~\ref{fig:hm-coverage1}). We refer to Section~\ref{sec:num_comparison} for further discussion. The analytical results are computed with $100$ terms, remarkably taking less than $5$~s to generate the whole plot. For comparison, the expression in~\cite{Mil08} provides the same output in about $248$~s.

Fig.~\ref{fig:kappa-bep-pt} illustrates the \ac{BEP} for coherent \ac{BPSK} versus the transmit power, with $\alpha\in \{0,0.3\}$.
For $P_\textrm{t} = 30$~dBm, doubling the number of antennas from $N = 256$ to $N = 512$ results in a $136 \times$ reduction in the \ac{BEP} with $\alpha = 0$, i.e., from $1.2 \times 10^{-3}$ to $8.8 \times 10^{-6}$. The analytical results are computed with no more than $400$ terms and the entire plot takes less than $3$~s to generate.

\begin{figure}[t]
    \centering
    \includegraphics[]{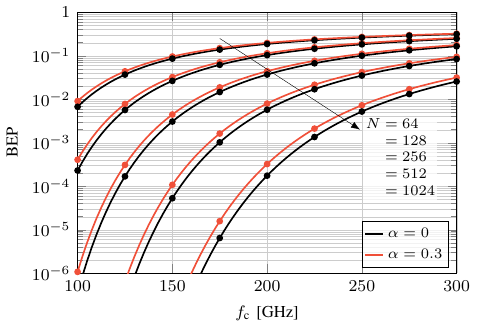}
    \caption{\km{} model: Downlink \ac{BEP} for coherent \ac{BPSK} versus carrier frequency, with $\kappa = 0.5$, 2$\mu = 0.5$, $d = 300$~m, $P_{\textrm{t}} = 30$~dBm, $N \in \{64,128,256,512,1024\}$, and $\alpha \in\{0,0.3\}$.}
    \label{fig:kappa-bep-freq}
\end{figure}

Fig.~\ref{fig:kappa-bep-freq} depicts the \ac{BEP} for coherent \ac{BPSK} versus the carrier frequency, again with $\alpha\in\{0, 0.3\}$.
Fixing a target \ac{BEP} of $10^{-3}$, increasing the number of antennas from $N=128$ to $N=1024$ allows to expand the carrier frequency approximately from $109$~GHz to $220$~GHz and the corresponding bandwidth from about $1.09$~GHz to $2.2$~GHz.

\begin{figure}[t]
    \centering
    \includegraphics[]{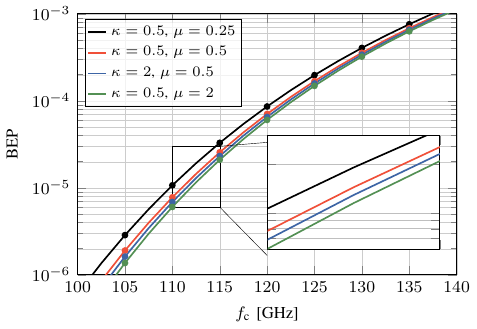}
    \caption{\km{} model: Downlink \ac{BEP} for coherent \ac{BPSK} versus carrier frequency, with $\kappa \in \{0.5, 2\}$, $\mu \in \{0.25, 0.5, 2\}$, $d = 300$~m, $P_{\textrm{t}} = 30$~dBm, $N = 256$, and $\alpha=0$, .}
    \label{fig:kappa-bep3}
\end{figure}

Fig.~\ref{fig:kappa-bep3} shows the \ac{BEP} versus the carrier frequency, with $N = 256$. We consider four combinations of fading parameters mimicking fading conditions ranging from very severe to favorable: (i) $\kappa=0.5$, $\mu=0.25$ (black line); (ii) $\kappa=0.5$, $\mu=0.5$ (red line); (iii) $\kappa=2$, $\mu=0.5$ (blue line); and (iv) $\kappa=0.5$, $\mu=2$ (green line). In (i), the \ac{BEP} is undermined by the small value of $\mu$ (i.e., the number of multipath clusters). In (ii), by slightly increasing $\mu$, the \ac{BEP} significantly improves, which allows to extend the carrier frequency by at least $1$~GHz. In (iii), a $4 \times$ increase in $\kappa$ (i.e., the power ratio between the \ac{LoS} and \ac{NLoS} components) produces only a timid decrease in the \ac{BEP} since the array gain weakly depends on $\kappa$. Lastly, in (vi), a bigger increase in $\mu$ yields a considerable \ac{BEP} improvement, further pushing the carrier frequency up by $1$~GHz. Hence, even though we consider \ac{LoS}-dominated propagation, the \ac{NLoS} components still have a noticeable impact on the system's performance.

\begin{figure}[t]
    \centering
    \includegraphics[]{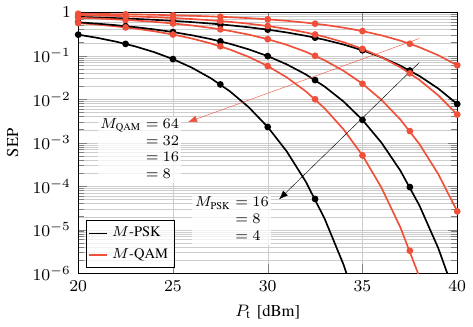}
    \caption{\km{} model: Downlink \ac{SEP} for \ac{$M$-PSK} and \ac{$M$-QAM} versus transmit power, with $\kappa = 0.5$, $\mu = 0.5$, $f_{\textrm{c}} = 140$~GHz, $d = 300$~m, $N = 512$, $\alpha=0$, $M_{\textrm{PSK}} \in \{4,8,16 \}$, and $M_{\textrm{QAM}} \in \{8,16,32,64\}$.}
    \label{fig:kappa-sep1}
\end{figure}

Fig.~\ref{fig:kappa-sep1} plots the \ac{SEP} for \ac{$M$-PSK} and \ac{$M$-QAM} versus the transmit power, with $N=512$, $M_{\textrm{PSK}} \in \{4,8,16 \}$, and $M_{\textrm{QAM}} \in \{8,16,32,64\}$. In this setting, the \ac{SEP} is lower than that of the FR3 system in Section~\ref{sec:num_mmwave} (cf. Fig.~\ref{fig:eta-sep1}): for instance, for $P_{\textrm{t}} = 30$~dBm and $4$-PSK, the \ac{SEP} decreases from $7.5 \times 10^{-3}$ to $1.4 \times 10^{-3}$.

\begin{figure}[t]
    \centering
    \includegraphics[]{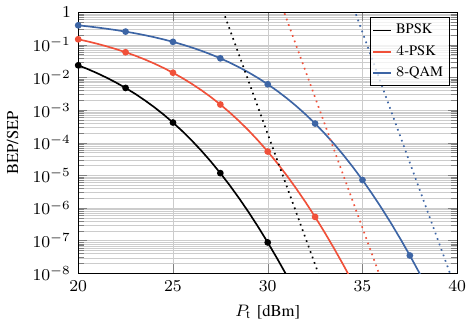}
    \caption{\km{} model: Downlink \ac{BEP} for coherent \ac{BPSK} and \ac{SEP} for $4$-PSK and $8$-QAM versus transmit power, with $\kappa = 0.5$, $\mu = 0.5$, $f_{\textrm{c}} = 140$~GHz, $d = 50$~m, $N = 32$, and $\alpha=0$.}
    \label{fig:kappa-asympt}
\end{figure}

Lastly, Fig.~\ref{fig:kappa-asympt} illustrates the asymptotic \ac{BEP} for coherent \ac{BPSK} and the asymptotic \ac{SEP} for $4$-PSK and $8$-QAM at high SNR versus the transmit power. To visualize the asymptotic behavior (dotted lines), we reduce the distance and the number of antennas to $d=50$ and $N=32$, respectively. As in Fig.~\ref{fig:eta-asympt}, the slope is mostly determined by the number of antennas and number of multipath clusters.

\subsection{Comparison Between FR3 and Sub-THz Systems}\label{sec:num_comparison}

Based on the results in Sections~\ref{sec:num_mmwave} and~\ref{sec:num_subthz}, we highlight that the number of antennas $N$, the number of multipath clusters $\mu$, and the path loss exponent $\beta$ are the key parameters that most significantly influence the system's performance. In both FR3 and sub-THz systems, $N$ and $\mu$ are the primary factors contributing to the array gain. We observe that an eight-fold increase in the number of antennas enables the carrier frequency (and, consequently, the bandwidth) to be approximately doubled while maintaining the same performance for both the FR3 and sub-THz systems considered. In addition, even in high-frequencies, \ac{LoS}-dominated scenarios, the \ac{NLoS} components continue to make a noticeable contribution to the received signal. Lastly, the \ac{LoS} or \ac{NLoS} path loss exponent has a greater impact on the system's performance than the frequency-dependent path loss factor. For example, compare the coverage probabilities in Fig.~\ref{fig:hm-coverage1} and Fig.~\ref{fig:kappa-coverage2}. Considering $d = 500$~m, the path loss at $f_{\textrm{c}} = 15$~GHz with $\beta = 3$ is about $5.74 \times$ stronger than at $f_{\textrm{c}} = 140$~GHz with $\beta = 2$. This is because the \ac{NLoS}-dominated propagation considered for the FR3 system is more detrimental than the increase in carrier frequency from $15$~GHz to $140$~GHz. Hence, for the same number of antennas, we observe a better coverage with \ac{LoS}-dominated propagation at $f_{\textrm{c}} = 140$~GHz than with \ac{NLoS}-dominated propagation at $f_{\textrm{c}} = 15$~GHz.

\section{Final Remarks} \label{sec:final}
In this paper, we developed a new exact representation of the sum of squared \ac{i.i.d.} \Ehm{} and \km{} \acp{RV}. The proposed analytical framework is remarkably tractable and computationally efficient, and thus can be conveniently employed to analyze system with massive antenna arrays. We derived novel expressions for the \ac{PDF} and \ac{CDF}, we analyzed their convergence and truncation error, and we discussed the computational complexity and implementation aspects. Furthermore, we derived expressions for the outage and coverage probability, \ac{BEP} for coherent binary modulations, and \ac{SEP} for \ac{$M$-PSK} and \ac{$M$-QAM}. Lastly, we provided an extensive performance evaluation of FR3 and sub-THz systems focusing on a downlink scenario where a single-antenna user is served by a \ac{BS} employing \ac{MRT}. The results unveiled the relationship between the system's performance and parameters such as the number of antennas, the number of multipath clusters, and the path loss exponent. Future work will consider extensions to the multi-user scenario.

\bibliographystyle{IEEEtran}
\bibliography{refs_abbr,refs}

\end{document}